%% file: main.tex
\documentclass[sigconf]{acmart}

\usepackage{xspace}

\AtBeginDocument{%
  \providecommand\BibTeX{{%
    \normalfont B\kern-0.5em{\scshape i\kern-0.25em b}\kern-0.8em\TeX}}}

\copyrightyear{2021}
\acmYear{2021}
\setcopyright{rightsretained}
\acmConference[SC '21]{The International Conference for High Performance Computing, Networking, Storage and Analysis}{November 14--19, 2021}{St. Louis, MO, USA}
\acmBooktitle{The International Conference for High Performance Computing, Networking, Storage and Analysis (SC '21), November 14--19, 2021, St. Louis, MO, USA}
\acmDOI{10.1145/3458817.3476184}
\acmISBN{978-1-4503-8442-1/21/11}

\input{macros.tex}

\input{color_macros.tex}
\input{eval_macros.tex}

\begin{document}

\title{\ourtitleSplit}

\author{Jack Kosaian}
\affiliation{%
  \institution{Carnegie Mellon University}
  \streetaddress{}
  \city{}
  \country{}
  }
\email{jkosaian@cs.cmu.edu}

\author{K.~V. Rashmi}
\affiliation{%
  \institution{Carnegie Mellon University}
  \streetaddress{}
  \city{}
  \country{}
  }
\email{rvinayak@cs.cmu.edu}

\renewcommand{\shortauthors}{Kosaian and Rashmi}
\renewcommand{\shorttitle}{\ourtitle}

\input{abstract.tex}



\maketitle

\input{intro.tex}
\input{background.tex}
\input{opportunity.tex}
\input{technique.tex}
\input{evaluation.tex}
\input{discussion.tex}
\input{related.tex}
\input{conclusion.tex}

{\footnotesize \bibliographystyle{acm}
\bibliography{references}}

\end{document}

%% file: macros.tex
\newcommand{\jack}[1]{{\color{red}[Jack: #1]}}
\newcommand{\rashmi}[1]{{\color{blue}[Rashmi: #1]}}
\newcommand{\rgood}{{\color{blue}[Rashmi: Looks good]}}

\newcommand{\rrem}[1]{{\color{cyan}[Rashmi removed: #1]}}
\newcommand{\updated}[1]{{\color{violet}[New: #1]}}
\newcommand{\remove}[1]{{\color{cyan}[Remove: #1]}}
\newcommand{\cutcandidate}[1]{{\color{cyan}[Cut? #1]}}
\newcommand{\revised}[1]{{\color{red}#1}}
\newcommand{\revisedremove}[1]{{\color{red}\st{#1}}}

\renewcommand{\jack}[1]{}
\renewcommand{\rashmi}[1]{}
\renewcommand{\updated}[1]{#1}
\renewcommand{\remove}[1]{}
\renewcommand{\cutcandidate}[1]{}

\renewcommand{\rrem}[1]{} 
\renewcommand{\rgood}[1]{}
\renewcommand{\revised}[1]{#1}
\renewcommand{\revisedremove}[1]{}

\newcommand{\ourtitleSplit}{Arithmetic-Intensity-Guided Fault Tolerance for\\Neural Network Inference on GPUs}
\newcommand{\ourtitle}{Arithmetic-Intensity-Guided Fault Tolerance for Neural Network Inference on GPUs}

\usepackage{enumitem}

\newenvironment{denseitemize}{
\begin{itemize}[topsep=2.5pt, partopsep=0pt, leftmargin=1.5em]
  \setlength{\itemsep}{2.5pt}
  \setlength{\parskip}{0pt}
  \setlength{\parsep}{0pt}
}{\end{itemize}}

\newcommand{\Section}{\S}
\newcommand{\Figure}{Figure~}
\newcommand{\Figures}{Figures~}
\newcommand{\eg}{e.g.,\xspace}

\newcommand{\cmr}{CMR\xspace}
\newcommand{\cmrs}{CMRs\xspace}
\newcommand{\cmrFull}{compute-to-memory-bandwidth ratio\xspace}

\newcommand{\FLOPS}{FLOPs/sec\xspace}
\newcommand{\TFLOPS}{TFLOPs/sec\xspace}
\newcommand{\TOPS}{TOPs/sec\xspace}
\newcommand{\ops}{FLOPs\xspace}

\newcommand{\nnShort}{NN\xspace}
\newcommand{\nnsShort}{NNs\xspace}

\newcommand{\NnsShort}{NNs\xspace}

\newcommand{\nn}{\nnShort}
\newcommand{\nns}{\nnsShort}

\newcommand{\Nns}{\NnsShort}

\newcommand{\smsLong}{streaming multiprocessors\xspace}

\newcommand{\sm}{SM\xspace}
\newcommand{\sms}{SMs\xspace}

\newcommand{\tFourHalfTFLOPS}{65\xspace}
\newcommand{\tFourGBps}{320\xspace}
\newcommand{\tFourHalfCMR}{203\xspace}
\newcommand{\pFourHalfCMR}{58\xspace}

\newcommand{\kernel}{kernel\xspace}
\newcommand{\kernels}{kernels\xspace}

\newcommand{\threadblock}{threadblock\xspace}
\newcommand{\threadblocks}{threadblocks\xspace}

\newcommand{\warp}{warp\xspace}
\newcommand{\warps}{warps\xspace}

\newcommand{\tensorcore}{Tensor Core\xspace}
\newcommand{\tensorcores}{Tensor Cores\xspace}

\newcommand{\dlrm}{DLRM\xspace}
\newcommand{\dlrms}{DLRMs\xspace}
\newcommand{\DLRM}{DLRM\xspace}

\newcommand{\matmatmultShort}{matrix multiplication\xspace}
\newcommand{\matmatmultsShort}{matrix multiplications\xspace}

\newcommand{\matmatmult}{matrix multiplication\xspace}
\newcommand{\matmatmults}{matrix multiplications\xspace}
\newcommand{\Matmatmult}{Matrix multiplication\xspace}

\newcommand{\matmult}{matrix multiplication\xspace}

\newcommand{\checksumA}{column checksum\xspace}
\newcommand{\checksumAs}{column checksums\xspace}

\newcommand{\checksumB}{row checksum\xspace}

\newcommand{\checksumC}{output summation\xspace}

\newcommand{\actChecksum}{activation checksum\xspace}

\newcommand{\weightChecksum}{weight checksum\xspace}
\newcommand{\weightChecksums}{weight checksums\xspace}

\newcommand{\cnn}{CNN\xspace}
\newcommand{\cnns}{CNNs\xspace}

\newcommand{\abft}{ABFT\xspace}
\newcommand{\threadABFT}{thread-level \abft}
\newcommand{\ThreadABFT}{Thread-level \abft}
\newcommand{\globalABFT}{global \abft}
\newcommand{\GlobalABFT}{Global \abft}
\newcommand{\hybridABFT}{intensity-guided ABFT\xspace}
\newcommand{\HybridABFT}{Intensity-guided ABFT\xspace}

\newcommand{\mSixteenNeightKeight}{m16n8k8\xspace}
\newcommand{\cutlass}{CUTLASS\xspace}

\newcommand{\matA}{A\xspace}
\newcommand{\matB}{B\xspace}
\newcommand{\matC}{C\xspace}
\newcommand{\matAtc}{A_{tc}\xspace}
\newcommand{\matBtc}{B_{tc}\xspace}
\newcommand{\matCtc}{C_{tc}\xspace}
\newcommand{\matAThread}{A_t\xspace}
\newcommand{\matBThread}{B_t\xspace}
\newcommand{\mma}{MMA\xspace}
\newcommand{\mmas}{MMAs\xspace}
\newcommand{\gemmM}{M\xspace}
\newcommand{\gemmN}{N\xspace}
\newcommand{\gemmK}{K\xspace}
\newcommand{\gemmMThread}{M_t\xspace}
\newcommand{\gemmNThread}{N_t\xspace}

\newcommand{\ai}{arithmetic intensity\xspace}

\newcommand{\aiShort}{arithmetic intensity\xspace}
\newcommand{\aisShort}{arithmetic intensities\xspace}
\newcommand{\aggregateAI}{aggregate \aiShort}

\newcommand{\aggregateAIs}{aggregate \aisShort}

\newcommand{\se}{soft error\xspace}
\newcommand{\ses}{soft errors\xspace}

\newcommand{\SlowdownMetric}{Time overhead\xspace}

\newcommand{\harietal}{Hari et al.~\cite{hari2021making}\xspace}

\newcommand{\nameAlexnet}{AlexNet\xspace}
\newcommand{\nameDensenet}{DenseNet-161\xspace}
\newcommand{\nameResnet}{ResNet-50\xspace}
\newcommand{\nameResnext}{ResNext-50\xspace}
\newcommand{\nameShufflenet}{ShuffleNet\xspace}
\newcommand{\nameSqueezenet}{SqueezeNet\xspace}
\newcommand{\nameVgg}{VGG-16\xspace}
\newcommand{\nameWideResnet}{Wide-ResNet-50\xspace}
\newcommand{\nameWideResnetShort}{WideResNet\xspace}

\newcommand{\nameDLRMBottom}{MLP-Bottom\xspace}
\newcommand{\nameDLRMTop}{MLP-Top\xspace}

\newcommand{\shortNameNoscopeCoral}{Coral\xspace}
\newcommand{\shortNameNoscopeRoundabout}{Roundabout\xspace}
\newcommand{\shortNameNoscopeTaipei}{Taipei\xspace}
\newcommand{\shortNameNoscopeAmsterdam}{Amsterdam\xspace}

\newcommand{\numThreadMMA}{$\frac{\gemmMThread\gemmNThread}{2}$\xspace}
\newcommand{\numThreadMMABold}{$\mathbf{\frac{\gemmMThread\gemmNThread}{2}}$\xspace}
\newcommand{\numThreadOutReg}{$\gemmMThread\gemmNThread$\xspace}

\newcommand{\linearlayer}{linear layer\xspace}
\newcommand{\linearlayers}{linear layers\xspace}

\newcommand{\layer}{layer\xspace}
\newcommand{\layers}{layers\xspace}

%% file: color_macros.tex
\usepackage{tikz}
\usetikzlibrary{spy}
\usetikzlibrary{patterns}
\usepackage{pgfplots}
\definecolor{color_black_2_0}{rgb}{0.9411764705882353,0.9411764705882353,0.9411764705882353}
\definecolor{color_black_2_1}{rgb}{0.38823529411764707,0.38823529411764707,0.38823529411764707}
\definecolor{color_black_3_0}{rgb}{0.9411764705882353,0.9411764705882353,0.9411764705882353}
\definecolor{color_black_3_1}{rgb}{0.7411764705882353,0.7411764705882353,0.7411764705882353}
\definecolor{color_black_3_2}{rgb}{0.38823529411764707,0.38823529411764707,0.38823529411764707}
\definecolor{color_black_4_0}{rgb}{0.9686274509803922,0.9686274509803922,0.9686274509803922}
\definecolor{color_black_4_1}{rgb}{0.8,0.8,0.8}
\definecolor{color_black_4_2}{rgb}{0.5882352941176471,0.5882352941176471,0.5882352941176471}
\definecolor{color_black_4_3}{rgb}{0.3215686274509804,0.3215686274509804,0.3215686274509804}
\definecolor{color_black_5_0}{rgb}{0.9686274509803922,0.9686274509803922,0.9686274509803922}
\definecolor{color_black_5_1}{rgb}{0.8,0.8,0.8}
\definecolor{color_black_5_2}{rgb}{0.5882352941176471,0.5882352941176471,0.5882352941176471}
\definecolor{color_black_5_3}{rgb}{0.38823529411764707,0.38823529411764707,0.38823529411764707}
\definecolor{color_black_5_4}{rgb}{0.1450980392156863,0.1450980392156863,0.1450980392156863}
\definecolor{color_black_6_0}{rgb}{0.9686274509803922,0.9686274509803922,0.9686274509803922}
\definecolor{color_black_6_1}{rgb}{0.8509803921568627,0.8509803921568627,0.8509803921568627}
\definecolor{color_black_6_2}{rgb}{0.7411764705882353,0.7411764705882353,0.7411764705882353}
\definecolor{color_black_6_3}{rgb}{0.5882352941176471,0.5882352941176471,0.5882352941176471}
\definecolor{color_black_6_4}{rgb}{0.38823529411764707,0.38823529411764707,0.38823529411764707}
\definecolor{color_black_6_5}{rgb}{0.1450980392156863,0.1450980392156863,0.1450980392156863}
\definecolor{color_black_7_0}{rgb}{0.9686274509803922,0.9686274509803922,0.9686274509803922}
\definecolor{color_black_7_1}{rgb}{0.8509803921568627,0.8509803921568627,0.8509803921568627}
\definecolor{color_black_7_2}{rgb}{0.7411764705882353,0.7411764705882353,0.7411764705882353}
\definecolor{color_black_7_3}{rgb}{0.5882352941176471,0.5882352941176471,0.5882352941176471}
\definecolor{color_black_7_4}{rgb}{0.45098039215686275,0.45098039215686275,0.45098039215686275}
\definecolor{color_black_7_5}{rgb}{0.3215686274509804,0.3215686274509804,0.3215686274509804}
\definecolor{color_black_7_6}{rgb}{0.1450980392156863,0.1450980392156863,0.1450980392156863}
\definecolor{color_black_8_0}{rgb}{1.0,1.0,1.0}
\definecolor{color_black_8_1}{rgb}{0.9411764705882353,0.9411764705882353,0.9411764705882353}
\definecolor{color_black_8_2}{rgb}{0.8509803921568627,0.8509803921568627,0.8509803921568627}
\definecolor{color_black_8_3}{rgb}{0.7411764705882353,0.7411764705882353,0.7411764705882353}
\definecolor{color_black_8_4}{rgb}{0.5882352941176471,0.5882352941176471,0.5882352941176471}
\definecolor{color_black_8_5}{rgb}{0.45098039215686275,0.45098039215686275,0.45098039215686275}
\definecolor{color_black_8_6}{rgb}{0.3215686274509804,0.3215686274509804,0.3215686274509804}
\definecolor{color_black_8_7}{rgb}{0.1450980392156863,0.1450980392156863,0.1450980392156863}
\definecolor{color_black_9_0}{rgb}{1.0,1.0,1.0}
\definecolor{color_black_9_1}{rgb}{0.9411764705882353,0.9411764705882353,0.9411764705882353}
\definecolor{color_black_9_2}{rgb}{0.8509803921568627,0.8509803921568627,0.8509803921568627}
\definecolor{color_black_9_3}{rgb}{0.7411764705882353,0.7411764705882353,0.7411764705882353}
\definecolor{color_black_9_4}{rgb}{0.5882352941176471,0.5882352941176471,0.5882352941176471}
\definecolor{color_black_9_5}{rgb}{0.45098039215686275,0.45098039215686275,0.45098039215686275}
\definecolor{color_black_9_6}{rgb}{0.3215686274509804,0.3215686274509804,0.3215686274509804}
\definecolor{color_black_9_7}{rgb}{0.1450980392156863,0.1450980392156863,0.1450980392156863}
\definecolor{color_black_9_8}{rgb}{0.0,0.0,0.0}
\definecolor{color_blue_2_0}{rgb}{0.8705882352941177,0.9215686274509803,0.9686274509803922}
\definecolor{color_blue_2_1}{rgb}{0.19215686274509805,0.5098039215686274,0.7411764705882353}
\definecolor{color_blue_3_0}{rgb}{0.8705882352941177,0.9215686274509803,0.9686274509803922}
\definecolor{color_blue_3_1}{rgb}{0.6196078431372549,0.792156862745098,0.8823529411764706}
\definecolor{color_blue_3_2}{rgb}{0.19215686274509805,0.5098039215686274,0.7411764705882353}
\definecolor{color_blue_4_0}{rgb}{0.9372549019607843,0.9529411764705882,1.0}
\definecolor{color_blue_4_1}{rgb}{0.7411764705882353,0.8431372549019608,0.9058823529411765}
\definecolor{color_blue_4_2}{rgb}{0.4196078431372549,0.6823529411764706,0.8392156862745098}
\definecolor{color_blue_4_3}{rgb}{0.12941176470588237,0.44313725490196076,0.7098039215686275}
\definecolor{color_blue_5_0}{rgb}{0.9372549019607843,0.9529411764705882,1.0}
\definecolor{color_blue_5_1}{rgb}{0.7411764705882353,0.8431372549019608,0.9058823529411765}
\definecolor{color_blue_5_2}{rgb}{0.4196078431372549,0.6823529411764706,0.8392156862745098}
\definecolor{color_blue_5_3}{rgb}{0.19215686274509805,0.5098039215686274,0.7411764705882353}
\definecolor{color_blue_5_4}{rgb}{0.03137254901960784,0.3176470588235294,0.611764705882353}
\definecolor{color_blue_6_0}{rgb}{0.9372549019607843,0.9529411764705882,1.0}
\definecolor{color_blue_6_1}{rgb}{0.7764705882352941,0.8588235294117647,0.9372549019607843}
\definecolor{color_blue_6_2}{rgb}{0.6196078431372549,0.792156862745098,0.8823529411764706}
\definecolor{color_blue_6_3}{rgb}{0.4196078431372549,0.6823529411764706,0.8392156862745098}
\definecolor{color_blue_6_4}{rgb}{0.19215686274509805,0.5098039215686274,0.7411764705882353}
\definecolor{color_blue_6_5}{rgb}{0.03137254901960784,0.3176470588235294,0.611764705882353}
\definecolor{color_blue_7_0}{rgb}{0.9372549019607843,0.9529411764705882,1.0}
\definecolor{color_blue_7_1}{rgb}{0.7764705882352941,0.8588235294117647,0.9372549019607843}
\definecolor{color_blue_7_2}{rgb}{0.6196078431372549,0.792156862745098,0.8823529411764706}
\definecolor{color_blue_7_3}{rgb}{0.4196078431372549,0.6823529411764706,0.8392156862745098}
\definecolor{color_blue_7_4}{rgb}{0.25882352941176473,0.5725490196078431,0.7764705882352941}
\definecolor{color_blue_7_5}{rgb}{0.12941176470588237,0.44313725490196076,0.7098039215686275}
\definecolor{color_blue_7_6}{rgb}{0.03137254901960784,0.27058823529411763,0.5803921568627451}
\definecolor{color_blue_8_0}{rgb}{0.9686274509803922,0.984313725490196,1.0}
\definecolor{color_blue_8_1}{rgb}{0.8705882352941177,0.9215686274509803,0.9686274509803922}
\definecolor{color_blue_8_2}{rgb}{0.7764705882352941,0.8588235294117647,0.9372549019607843}
\definecolor{color_blue_8_3}{rgb}{0.6196078431372549,0.792156862745098,0.8823529411764706}
\definecolor{color_blue_8_4}{rgb}{0.4196078431372549,0.6823529411764706,0.8392156862745098}
\definecolor{color_blue_8_5}{rgb}{0.25882352941176473,0.5725490196078431,0.7764705882352941}
\definecolor{color_blue_8_6}{rgb}{0.12941176470588237,0.44313725490196076,0.7098039215686275}
\definecolor{color_blue_8_7}{rgb}{0.03137254901960784,0.27058823529411763,0.5803921568627451}
\definecolor{color_blue_9_0}{rgb}{0.9686274509803922,0.984313725490196,1.0}
\definecolor{color_blue_9_1}{rgb}{0.8705882352941177,0.9215686274509803,0.9686274509803922}
\definecolor{color_blue_9_2}{rgb}{0.7764705882352941,0.8588235294117647,0.9372549019607843}
\definecolor{color_blue_9_3}{rgb}{0.6196078431372549,0.792156862745098,0.8823529411764706}
\definecolor{color_blue_9_4}{rgb}{0.4196078431372549,0.6823529411764706,0.8392156862745098}
\definecolor{color_blue_9_5}{rgb}{0.25882352941176473,0.5725490196078431,0.7764705882352941}
\definecolor{color_blue_9_6}{rgb}{0.12941176470588237,0.44313725490196076,0.7098039215686275}
\definecolor{color_blue_9_7}{rgb}{0.03137254901960784,0.3176470588235294,0.611764705882353}
\definecolor{color_blue_9_8}{rgb}{0.03137254901960784,0.18823529411764706,0.4196078431372549}
\definecolor{color_green_2_0}{rgb}{0.8980392156862745,0.9607843137254902,0.8784313725490196}
\definecolor{color_green_2_1}{rgb}{0.19215686274509805,0.6392156862745098,0.32941176470588235}
\definecolor{color_green_3_0}{rgb}{0.8980392156862745,0.9607843137254902,0.8784313725490196}
\definecolor{color_green_3_1}{rgb}{0.6313725490196078,0.8509803921568627,0.6078431372549019}
\definecolor{color_green_3_2}{rgb}{0.19215686274509805,0.6392156862745098,0.32941176470588235}
\definecolor{color_green_4_0}{rgb}{0.9294117647058824,0.9725490196078431,0.9137254901960784}
\definecolor{color_green_4_1}{rgb}{0.7294117647058823,0.8941176470588236,0.7019607843137254}
\definecolor{color_green_4_2}{rgb}{0.4549019607843137,0.7686274509803922,0.4627450980392157}
\definecolor{color_green_4_3}{rgb}{0.13725490196078433,0.5450980392156862,0.27058823529411763}
\definecolor{color_green_5_0}{rgb}{0.9294117647058824,0.9725490196078431,0.9137254901960784}
\definecolor{color_green_5_1}{rgb}{0.7294117647058823,0.8941176470588236,0.7019607843137254}
\definecolor{color_green_5_2}{rgb}{0.4549019607843137,0.7686274509803922,0.4627450980392157}
\definecolor{color_green_5_3}{rgb}{0.19215686274509805,0.6392156862745098,0.32941176470588235}
\definecolor{color_green_5_4}{rgb}{0.0,0.42745098039215684,0.17254901960784313}
\definecolor{color_green_6_0}{rgb}{0.9294117647058824,0.9725490196078431,0.9137254901960784}
\definecolor{color_green_6_1}{rgb}{0.7803921568627451,0.9137254901960784,0.7529411764705882}
\definecolor{color_green_6_2}{rgb}{0.6313725490196078,0.8509803921568627,0.6078431372549019}
\definecolor{color_green_6_3}{rgb}{0.4549019607843137,0.7686274509803922,0.4627450980392157}
\definecolor{color_green_6_4}{rgb}{0.19215686274509805,0.6392156862745098,0.32941176470588235}
\definecolor{color_green_6_5}{rgb}{0.0,0.42745098039215684,0.17254901960784313}
\definecolor{color_green_7_0}{rgb}{0.9294117647058824,0.9725490196078431,0.9137254901960784}
\definecolor{color_green_7_1}{rgb}{0.7803921568627451,0.9137254901960784,0.7529411764705882}
\definecolor{color_green_7_2}{rgb}{0.6313725490196078,0.8509803921568627,0.6078431372549019}
\definecolor{color_green_7_3}{rgb}{0.4549019607843137,0.7686274509803922,0.4627450980392157}
\definecolor{color_green_7_4}{rgb}{0.2549019607843137,0.6705882352941176,0.36470588235294116}
\definecolor{color_green_7_5}{rgb}{0.13725490196078433,0.5450980392156862,0.27058823529411763}
\definecolor{color_green_7_6}{rgb}{0.0,0.35294117647058826,0.19607843137254902}
\definecolor{color_green_8_0}{rgb}{0.9686274509803922,0.9882352941176471,0.9607843137254902}
\definecolor{color_green_8_1}{rgb}{0.8980392156862745,0.9607843137254902,0.8784313725490196}
\definecolor{color_green_8_2}{rgb}{0.7803921568627451,0.9137254901960784,0.7529411764705882}
\definecolor{color_green_8_3}{rgb}{0.6313725490196078,0.8509803921568627,0.6078431372549019}
\definecolor{color_green_8_4}{rgb}{0.4549019607843137,0.7686274509803922,0.4627450980392157}
\definecolor{color_green_8_5}{rgb}{0.2549019607843137,0.6705882352941176,0.36470588235294116}
\definecolor{color_green_8_6}{rgb}{0.13725490196078433,0.5450980392156862,0.27058823529411763}
\definecolor{color_green_8_7}{rgb}{0.0,0.35294117647058826,0.19607843137254902}
\definecolor{color_green_9_0}{rgb}{0.9686274509803922,0.9882352941176471,0.9607843137254902}
\definecolor{color_green_9_1}{rgb}{0.8980392156862745,0.9607843137254902,0.8784313725490196}
\definecolor{color_green_9_2}{rgb}{0.7803921568627451,0.9137254901960784,0.7529411764705882}
\definecolor{color_green_9_3}{rgb}{0.6313725490196078,0.8509803921568627,0.6078431372549019}
\definecolor{color_green_9_4}{rgb}{0.4549019607843137,0.7686274509803922,0.4627450980392157}
\definecolor{color_green_9_5}{rgb}{0.2549019607843137,0.6705882352941176,0.36470588235294116}
\definecolor{color_green_9_6}{rgb}{0.13725490196078433,0.5450980392156862,0.27058823529411763}
\definecolor{color_green_9_7}{rgb}{0.0,0.42745098039215684,0.17254901960784313}
\definecolor{color_green_9_8}{rgb}{0.0,0.26666666666666666,0.10588235294117647}
\definecolor{color_orange_2_0}{rgb}{0.996078431372549,0.9019607843137255,0.807843137254902}
\definecolor{color_orange_2_1}{rgb}{0.9019607843137255,0.3333333333333333,0.050980392156862744}
\definecolor{color_orange_3_0}{rgb}{0.996078431372549,0.9019607843137255,0.807843137254902}
\definecolor{color_orange_3_1}{rgb}{0.9921568627450981,0.6823529411764706,0.4196078431372549}
\definecolor{color_orange_3_2}{rgb}{0.9019607843137255,0.3333333333333333,0.050980392156862744}
\definecolor{color_orange_4_0}{rgb}{0.996078431372549,0.9294117647058824,0.8705882352941177}
\definecolor{color_orange_4_1}{rgb}{0.9921568627450981,0.7450980392156863,0.5215686274509804}
\definecolor{color_orange_4_2}{rgb}{0.9921568627450981,0.5529411764705883,0.23529411764705882}
\definecolor{color_orange_4_3}{rgb}{0.8509803921568627,0.2784313725490196,0.00392156862745098}
\definecolor{color_orange_5_0}{rgb}{0.996078431372549,0.9294117647058824,0.8705882352941177}
\definecolor{color_orange_5_1}{rgb}{0.9921568627450981,0.7450980392156863,0.5215686274509804}
\definecolor{color_orange_5_2}{rgb}{0.9921568627450981,0.5529411764705883,0.23529411764705882}
\definecolor{color_orange_5_3}{rgb}{0.9019607843137255,0.3333333333333333,0.050980392156862744}
\definecolor{color_orange_5_4}{rgb}{0.6509803921568628,0.21176470588235294,0.011764705882352941}
\definecolor{color_orange_6_0}{rgb}{0.996078431372549,0.9294117647058824,0.8705882352941177}
\definecolor{color_orange_6_1}{rgb}{0.9921568627450981,0.8156862745098039,0.6352941176470588}
\definecolor{color_orange_6_2}{rgb}{0.9921568627450981,0.6823529411764706,0.4196078431372549}
\definecolor{color_orange_6_3}{rgb}{0.9921568627450981,0.5529411764705883,0.23529411764705882}
\definecolor{color_orange_6_4}{rgb}{0.9019607843137255,0.3333333333333333,0.050980392156862744}
\definecolor{color_orange_6_5}{rgb}{0.6509803921568628,0.21176470588235294,0.011764705882352941}
\definecolor{color_orange_7_0}{rgb}{0.996078431372549,0.9294117647058824,0.8705882352941177}
\definecolor{color_orange_7_1}{rgb}{0.9921568627450981,0.8156862745098039,0.6352941176470588}
\definecolor{color_orange_7_2}{rgb}{0.9921568627450981,0.6823529411764706,0.4196078431372549}
\definecolor{color_orange_7_3}{rgb}{0.9921568627450981,0.5529411764705883,0.23529411764705882}
\definecolor{color_orange_7_4}{rgb}{0.9450980392156862,0.4117647058823529,0.07450980392156863}
\definecolor{color_orange_7_5}{rgb}{0.8509803921568627,0.2823529411764706,0.00392156862745098}
\definecolor{color_orange_7_6}{rgb}{0.5490196078431373,0.17647058823529413,0.01568627450980392}
\definecolor{color_orange_8_0}{rgb}{1.0,0.9607843137254902,0.9215686274509803}
\definecolor{color_orange_8_1}{rgb}{0.996078431372549,0.9019607843137255,0.807843137254902}
\definecolor{color_orange_8_2}{rgb}{0.9921568627450981,0.8156862745098039,0.6352941176470588}
\definecolor{color_orange_8_3}{rgb}{0.9921568627450981,0.6823529411764706,0.4196078431372549}
\definecolor{color_orange_8_4}{rgb}{0.9921568627450981,0.5529411764705883,0.23529411764705882}
\definecolor{color_orange_8_5}{rgb}{0.9450980392156862,0.4117647058823529,0.07450980392156863}
\definecolor{color_orange_8_6}{rgb}{0.8509803921568627,0.2823529411764706,0.00392156862745098}
\definecolor{color_orange_8_7}{rgb}{0.5490196078431373,0.17647058823529413,0.01568627450980392}
\definecolor{color_orange_9_0}{rgb}{1.0,0.9607843137254902,0.9215686274509803}
\definecolor{color_orange_9_1}{rgb}{0.996078431372549,0.9019607843137255,0.807843137254902}
\definecolor{color_orange_9_2}{rgb}{0.9921568627450981,0.8156862745098039,0.6352941176470588}
\definecolor{color_orange_9_3}{rgb}{0.9921568627450981,0.6823529411764706,0.4196078431372549}
\definecolor{color_orange_9_4}{rgb}{0.9921568627450981,0.5529411764705883,0.23529411764705882}
\definecolor{color_orange_9_5}{rgb}{0.9450980392156862,0.4117647058823529,0.07450980392156863}
\definecolor{color_orange_9_6}{rgb}{0.8509803921568627,0.2823529411764706,0.00392156862745098}
\definecolor{color_orange_9_7}{rgb}{0.6509803921568628,0.21176470588235294,0.011764705882352941}
\definecolor{color_orange_9_8}{rgb}{0.4980392156862745,0.15294117647058825,0.01568627450980392}
\definecolor{color_purple_2_0}{rgb}{0.9372549019607843,0.9294117647058824,0.9607843137254902}
\definecolor{color_purple_2_1}{rgb}{0.4588235294117647,0.4196078431372549,0.6941176470588235}
\definecolor{color_purple_3_0}{rgb}{0.9372549019607843,0.9294117647058824,0.9607843137254902}
\definecolor{color_purple_3_1}{rgb}{0.7372549019607844,0.7411764705882353,0.8627450980392157}
\definecolor{color_purple_3_2}{rgb}{0.4588235294117647,0.4196078431372549,0.6941176470588235}
\definecolor{color_purple_4_0}{rgb}{0.9490196078431372,0.9411764705882353,0.9686274509803922}
\definecolor{color_purple_4_1}{rgb}{0.796078431372549,0.788235294117647,0.8862745098039215}
\definecolor{color_purple_4_2}{rgb}{0.6196078431372549,0.6039215686274509,0.7843137254901961}
\definecolor{color_purple_4_3}{rgb}{0.41568627450980394,0.3176470588235294,0.6392156862745098}
\definecolor{color_purple_5_0}{rgb}{0.9490196078431372,0.9411764705882353,0.9686274509803922}
\definecolor{color_purple_5_1}{rgb}{0.796078431372549,0.788235294117647,0.8862745098039215}
\definecolor{color_purple_5_2}{rgb}{0.6196078431372549,0.6039215686274509,0.7843137254901961}
\definecolor{color_purple_5_3}{rgb}{0.4588235294117647,0.4196078431372549,0.6941176470588235}
\definecolor{color_purple_5_4}{rgb}{0.32941176470588235,0.15294117647058825,0.5607843137254902}
\definecolor{color_purple_6_0}{rgb}{0.9490196078431372,0.9411764705882353,0.9686274509803922}
\definecolor{color_purple_6_1}{rgb}{0.8549019607843137,0.8549019607843137,0.9215686274509803}
\definecolor{color_purple_6_2}{rgb}{0.7372549019607844,0.7411764705882353,0.8627450980392157}
\definecolor{color_purple_6_3}{rgb}{0.6196078431372549,0.6039215686274509,0.7843137254901961}
\definecolor{color_purple_6_4}{rgb}{0.4588235294117647,0.4196078431372549,0.6941176470588235}
\definecolor{color_purple_6_5}{rgb}{0.32941176470588235,0.15294117647058825,0.5607843137254902}
\definecolor{color_purple_7_0}{rgb}{0.9490196078431372,0.9411764705882353,0.9686274509803922}
\definecolor{color_purple_7_1}{rgb}{0.8549019607843137,0.8549019607843137,0.9215686274509803}
\definecolor{color_purple_7_2}{rgb}{0.7372549019607844,0.7411764705882353,0.8627450980392157}
\definecolor{color_purple_7_3}{rgb}{0.6196078431372549,0.6039215686274509,0.7843137254901961}
\definecolor{color_purple_7_4}{rgb}{0.5019607843137255,0.49019607843137253,0.7294117647058823}
\definecolor{color_purple_7_5}{rgb}{0.41568627450980394,0.3176470588235294,0.6392156862745098}
\definecolor{color_purple_7_6}{rgb}{0.2901960784313726,0.0784313725490196,0.5254901960784314}
\definecolor{color_purple_8_0}{rgb}{0.9882352941176471,0.984313725490196,0.9921568627450981}
\definecolor{color_purple_8_1}{rgb}{0.9372549019607843,0.9294117647058824,0.9607843137254902}
\definecolor{color_purple_8_2}{rgb}{0.8549019607843137,0.8549019607843137,0.9215686274509803}
\definecolor{color_purple_8_3}{rgb}{0.7372549019607844,0.7411764705882353,0.8627450980392157}
\definecolor{color_purple_8_4}{rgb}{0.6196078431372549,0.6039215686274509,0.7843137254901961}
\definecolor{color_purple_8_5}{rgb}{0.5019607843137255,0.49019607843137253,0.7294117647058823}
\definecolor{color_purple_8_6}{rgb}{0.41568627450980394,0.3176470588235294,0.6392156862745098}
\definecolor{color_purple_8_7}{rgb}{0.2901960784313726,0.0784313725490196,0.5254901960784314}
\definecolor{color_purple_9_0}{rgb}{0.9882352941176471,0.984313725490196,0.9921568627450981}
\definecolor{color_purple_9_1}{rgb}{0.9372549019607843,0.9294117647058824,0.9607843137254902}
\definecolor{color_purple_9_2}{rgb}{0.8549019607843137,0.8549019607843137,0.9215686274509803}
\definecolor{color_purple_9_3}{rgb}{0.7372549019607844,0.7411764705882353,0.8627450980392157}
\definecolor{color_purple_9_4}{rgb}{0.6196078431372549,0.6039215686274509,0.7843137254901961}
\definecolor{color_purple_9_5}{rgb}{0.5019607843137255,0.49019607843137253,0.7294117647058823}
\definecolor{color_purple_9_6}{rgb}{0.41568627450980394,0.3176470588235294,0.6392156862745098}
\definecolor{color_purple_9_7}{rgb}{0.32941176470588235,0.15294117647058825,0.5607843137254902}
\definecolor{color_purple_9_8}{rgb}{0.24705882352941178,0.0,0.49019607843137253}
\definecolor{color_red_2_0}{rgb}{0.996078431372549,0.8784313725490196,0.8235294117647058}
\definecolor{color_red_2_1}{rgb}{0.8705882352941177,0.17647058823529413,0.14901960784313725}
\definecolor{color_red_3_0}{rgb}{0.996078431372549,0.8784313725490196,0.8235294117647058}
\definecolor{color_red_3_1}{rgb}{0.9882352941176471,0.5725490196078431,0.4470588235294118}
\definecolor{color_red_3_2}{rgb}{0.8705882352941177,0.17647058823529413,0.14901960784313725}
\definecolor{color_red_4_0}{rgb}{0.996078431372549,0.8980392156862745,0.8509803921568627}
\definecolor{color_red_4_1}{rgb}{0.9882352941176471,0.6823529411764706,0.5686274509803921}
\definecolor{color_red_4_2}{rgb}{0.984313725490196,0.41568627450980394,0.2901960784313726}
\definecolor{color_red_4_3}{rgb}{0.796078431372549,0.09411764705882353,0.11372549019607843}
\definecolor{color_red_5_0}{rgb}{0.996078431372549,0.8980392156862745,0.8509803921568627}
\definecolor{color_red_5_1}{rgb}{0.9882352941176471,0.6823529411764706,0.5686274509803921}
\definecolor{color_red_5_2}{rgb}{0.984313725490196,0.41568627450980394,0.2901960784313726}
\definecolor{color_red_5_3}{rgb}{0.8705882352941177,0.17647058823529413,0.14901960784313725}
\definecolor{color_red_5_4}{rgb}{0.6470588235294118,0.058823529411764705,0.08235294117647059}
\definecolor{color_red_6_0}{rgb}{0.996078431372549,0.8980392156862745,0.8509803921568627}
\definecolor{color_red_6_1}{rgb}{0.9882352941176471,0.7333333333333333,0.6313725490196078}
\definecolor{color_red_6_2}{rgb}{0.9882352941176471,0.5725490196078431,0.4470588235294118}
\definecolor{color_red_6_3}{rgb}{0.984313725490196,0.41568627450980394,0.2901960784313726}
\definecolor{color_red_6_4}{rgb}{0.8705882352941177,0.17647058823529413,0.14901960784313725}
\definecolor{color_red_6_5}{rgb}{0.6470588235294118,0.058823529411764705,0.08235294117647059}
\definecolor{color_red_7_0}{rgb}{0.996078431372549,0.8980392156862745,0.8509803921568627}
\definecolor{color_red_7_1}{rgb}{0.9882352941176471,0.7333333333333333,0.6313725490196078}
\definecolor{color_red_7_2}{rgb}{0.9882352941176471,0.5725490196078431,0.4470588235294118}
\definecolor{color_red_7_3}{rgb}{0.984313725490196,0.41568627450980394,0.2901960784313726}
\definecolor{color_red_7_4}{rgb}{0.9372549019607843,0.23137254901960785,0.17254901960784313}
\definecolor{color_red_7_5}{rgb}{0.796078431372549,0.09411764705882353,0.11372549019607843}
\definecolor{color_red_7_6}{rgb}{0.6,0.0,0.050980392156862744}
\definecolor{color_red_8_0}{rgb}{1.0,0.9607843137254902,0.9411764705882353}
\definecolor{color_red_8_1}{rgb}{0.996078431372549,0.8784313725490196,0.8235294117647058}
\definecolor{color_red_8_2}{rgb}{0.9882352941176471,0.7333333333333333,0.6313725490196078}
\definecolor{color_red_8_3}{rgb}{0.9882352941176471,0.5725490196078431,0.4470588235294118}
\definecolor{color_red_8_4}{rgb}{0.984313725490196,0.41568627450980394,0.2901960784313726}
\definecolor{color_red_8_5}{rgb}{0.9372549019607843,0.23137254901960785,0.17254901960784313}
\definecolor{color_red_8_6}{rgb}{0.796078431372549,0.09411764705882353,0.11372549019607843}
\definecolor{color_red_8_7}{rgb}{0.6,0.0,0.050980392156862744}
\definecolor{color_red_9_0}{rgb}{1.0,0.9607843137254902,0.9411764705882353}
\definecolor{color_red_9_1}{rgb}{0.996078431372549,0.8784313725490196,0.8235294117647058}
\definecolor{color_red_9_2}{rgb}{0.9882352941176471,0.7333333333333333,0.6313725490196078}
\definecolor{color_red_9_3}{rgb}{0.9882352941176471,0.5725490196078431,0.4470588235294118}
\definecolor{color_red_9_4}{rgb}{0.984313725490196,0.41568627450980394,0.2901960784313726}
\definecolor{color_red_9_5}{rgb}{0.9372549019607843,0.23137254901960785,0.17254901960784313}
\definecolor{color_red_9_6}{rgb}{0.796078431372549,0.09411764705882353,0.11372549019607843}
\definecolor{color_red_9_7}{rgb}{0.6470588235294118,0.058823529411764705,0.08235294117647059}
\definecolor{color_red_9_8}{rgb}{0.403921568627451,0.0,0.050980392156862744}

%% file: eval_macros.tex
\newcommand{\colorHybrid}{color_blue_2_0}
\newcommand{\colorGlobal}{color_blue_2_1}
\newcommand{\colorOgHybrid}{color_blue_3_0}
\newcommand{\colorOgThread}{color_blue_3_2}
\newcommand{\colorOgGlobal}{color_blue_3_1}

\newcommand{\colorErrorBars}{red}%
\newcommand{\colorErrorBarsGlobal}{black}%

\newcommand{\aiBatchOneResBigAlexnet}{125.5}
\newcommand{\threadOverheadBatchOneResBigAlexnet}{8.21}
\newcommand{\globalOverheadBatchOneResBigAlexnet}{4.08} %
\newcommand{\hybridOverheadBatchOneResBigAlexnet}{3.46}

\newcommand{\aiBatchOneResBigDensenet}{79.0}
\newcommand{\threadOverheadBatchOneResBigDensenet}{8.24}
\newcommand{\globalOverheadBatchOneResBigDensenet}{4.94} %
\newcommand{\hybridOverheadBatchOneResBigDensenet}{3.41}

\newcommand{\aiBatchOneResBigResnet}{122.0}
\newcommand{\threadOverheadBatchOneResBigResnet}{7.13}
\newcommand{\globalOverheadBatchOneResBigResnet}{8.49} %
\newcommand{\hybridOverheadBatchOneResBigResnet}{4.21}

\newcommand{\aiBatchOneResBigResnext}{220.8}
\newcommand{\threadOverheadBatchOneResBigResnext}{11.66}
\newcommand{\globalOverheadBatchOneResBigResnext}{5.46} %
\newcommand{\hybridOverheadBatchOneResBigResnext}{3.91}

\newcommand{\aiBatchOneResBigShufflenet}{76.6}
\newcommand{\threadOverheadBatchOneResBigShufflenet}{5.79}
\newcommand{\globalOverheadBatchOneResBigShufflenet}{11.39}%
\newcommand{\hybridOverheadBatchOneResBigShufflenet}{4.14}

\newcommand{\aiBatchOneResBigSqueezenet}{71.1}
\newcommand{\threadOverheadBatchOneResBigSqueezenet}{5.36}
\newcommand{\globalOverheadBatchOneResBigSqueezenet}{10.15}%
\newcommand{\hybridOverheadBatchOneResBigSqueezenet}{4.32}

\newcommand{\aiBatchOneResBigVgg}{155.5}
\newcommand{\threadOverheadBatchOneResBigVgg}{14.37}
\newcommand{\globalOverheadBatchOneResBigVgg}{3.43}%
\newcommand{\hybridOverheadBatchOneResBigVgg}{3.15}

\newcommand{\aiBatchOneResBigWideResnet}{220.8}
\newcommand{\threadOverheadBatchOneResBigWideResnet}{11.75}
\newcommand{\globalOverheadBatchOneResBigWideResnet}{5.25}%
\newcommand{\hybridOverheadBatchOneResBigWideResnet}{3.52}

\newcommand{\aiBatchOneDLRMBottom}{7.4}
\newcommand{\threadOverheadBatchOneDLRMBottom}{5.16}
\newcommand{\globalOverheadBatchOneDLRMBottom}{23.50} %
\newcommand{\hybridOverheadBatchOneDLRMBottom}{5.16}
\newcommand{\aiBatchOneDLRMTop}{7.7}
\newcommand{\threadOverheadBatchOneDLRMTop}{5.61}
\newcommand{\globalOverheadBatchOneDLRMTop}{18.16} %
\newcommand{\hybridOverheadBatchOneDLRMTop}{5.61}
\newcommand{\aiBatchTwentyFourtyEightDLRMBottom}{92.0}
\newcommand{\threadOverheadBatchTwentyFourtyEightDLRMBottom}{8.59}
\newcommand{\globalOverheadBatchTwentyFourtyEightDLRMBottom}{16.00}
\newcommand{\hybridOverheadBatchTwentyFourtyEightDLRMBottom}{6.96}
\newcommand{\aiBatchTwentyFourtyEightDLRMTop}{175.8}
\newcommand{\threadOverheadBatchTwentyFourtyEightDLRMTop}{10.79}
\newcommand{\globalOverheadBatchTwentyFourtyEightDLRMTop}{11.23}
\newcommand{\hybridOverheadBatchTwentyFourtyEightDLRMTop}{6.92}

\newcommand{\aiBatchSixtyFourNoscopeAmsterdam}{52.7}
\newcommand{\threadOverheadBatchSixtyFourNoscopeAmsterdam}{4.73}
\newcommand{\globalOverheadBatchSixtyFourNoscopeAmsterdam}{7.50} %
\newcommand{\hybridOverheadBatchSixtyFourNoscopeAmsterdam}{4.73}
\newcommand{\aiBatchSixtyFourNoscopeCoral}{15.1}
\newcommand{\threadOverheadBatchSixtyFourNoscopeCoral}{4.60}
\newcommand{\globalOverheadBatchSixtyFourNoscopeCoral}{17.07} %
\newcommand{\hybridOverheadBatchSixtyFourNoscopeCoral}{4.60}
\newcommand{\aiBatchSixtyFourNoscopeRoundabout}{37.9}
\newcommand{\threadOverheadBatchSixtyFourNoscopeRoundabout}{1.82}
\newcommand{\globalOverheadBatchSixtyFourNoscopeRoundabout}{9.61} %
\newcommand{\hybridOverheadBatchSixtyFourNoscopeRoundabout}{1.82}
\newcommand{\aiBatchSixtyFourNoscopeTaipei}{51.9}
\newcommand{\threadOverheadBatchSixtyFourNoscopeTaipei}{3.29}
\newcommand{\globalOverheadBatchSixtyFourNoscopeTaipei}{6.54} %
\newcommand{\hybridOverheadBatchSixtyFourNoscopeTaipei}{3.29}

\newcommand{\threadOverheadBatchOneResBigAlexnetPZero}{7.5291}
\newcommand{\threadOverheadBatchOneResBigAlexnetPOneHundred}{9.2887}
\newcommand{\globalOverheadBatchOneResBigAlexnetPZero}{2.7524}
\newcommand{\globalOverheadBatchOneResBigAlexnetPOneHundred}{5.8853}
\newcommand{\hybridOverheadBatchOneResBigAlexnetPZero}{2.2849}
\newcommand{\hybridOverheadBatchOneResBigAlexnetPOneHundred}{5.2250}
\newcommand{\threadOverheadBatchOneResBigDensenetPZero}{5.4569}
\newcommand{\threadOverheadBatchOneResBigDensenetPOneHundred}{10.0228}
\newcommand{\globalOverheadBatchOneResBigDensenetPZero}{3.8571}
\newcommand{\globalOverheadBatchOneResBigDensenetPOneHundred}{7.7581}
\newcommand{\hybridOverheadBatchOneResBigDensenetPZero}{2.4128}
\newcommand{\hybridOverheadBatchOneResBigDensenetPOneHundred}{5.7413}
\newcommand{\threadOverheadBatchOneResBigResnetPZero}{6.3483}
\newcommand{\threadOverheadBatchOneResBigResnetPOneHundred}{7.8373}
\newcommand{\globalOverheadBatchOneResBigResnetPZero}{7.8195}
\newcommand{\globalOverheadBatchOneResBigResnetPOneHundred}{9.1435}
\newcommand{\hybridOverheadBatchOneResBigResnetPZero}{3.8111}
\newcommand{\hybridOverheadBatchOneResBigResnetPOneHundred}{4.5745}
\newcommand{\threadOverheadBatchOneResBigResnextPZero}{10.8427}
\newcommand{\threadOverheadBatchOneResBigResnextPOneHundred}{12.6470}
\newcommand{\globalOverheadBatchOneResBigResnextPZero}{4.3834}
\newcommand{\globalOverheadBatchOneResBigResnextPOneHundred}{6.0795}
\newcommand{\hybridOverheadBatchOneResBigResnextPZero}{2.8021}
\newcommand{\hybridOverheadBatchOneResBigResnextPOneHundred}{4.6382}
\newcommand{\threadOverheadBatchOneResBigShufflenetPZero}{5.4625}
\newcommand{\threadOverheadBatchOneResBigShufflenetPOneHundred}{6.3261}
\newcommand{\globalOverheadBatchOneResBigShufflenetPZero}{11.3098}
\newcommand{\globalOverheadBatchOneResBigShufflenetPOneHundred}{11.5343}
\newcommand{\hybridOverheadBatchOneResBigShufflenetPZero}{4.0078}
\newcommand{\hybridOverheadBatchOneResBigShufflenetPOneHundred}{4.2894}
\newcommand{\threadOverheadBatchOneResBigSqueezenetPZero}{4.7294}
\newcommand{\threadOverheadBatchOneResBigSqueezenetPOneHundred}{6.4731}
\newcommand{\globalOverheadBatchOneResBigSqueezenetPZero}{8.9519}
\newcommand{\globalOverheadBatchOneResBigSqueezenetPOneHundred}{10.5649}
\newcommand{\hybridOverheadBatchOneResBigSqueezenetPZero}{3.2152}
\newcommand{\hybridOverheadBatchOneResBigSqueezenetPOneHundred}{5.1155}
\newcommand{\threadOverheadBatchOneResBigVggPZero}{9.8762}
\newcommand{\threadOverheadBatchOneResBigVggPOneHundred}{17.0708}
\newcommand{\globalOverheadBatchOneResBigVggPZero}{2.3882}
\newcommand{\globalOverheadBatchOneResBigVggPOneHundred}{4.9293}
\newcommand{\hybridOverheadBatchOneResBigVggPZero}{2.0968}
\newcommand{\hybridOverheadBatchOneResBigVggPOneHundred}{4.6728}
\newcommand{\threadOverheadBatchOneResBigWideResnetPZero}{10.6749}
\newcommand{\threadOverheadBatchOneResBigWideResnetPOneHundred}{12.6424}
\newcommand{\globalOverheadBatchOneResBigWideResnetPZero}{4.3325}
\newcommand{\globalOverheadBatchOneResBigWideResnetPOneHundred}{5.9325}
\newcommand{\hybridOverheadBatchOneResBigWideResnetPZero}{2.6932}
\newcommand{\hybridOverheadBatchOneResBigWideResnetPOneHundred}{4.2170}

\newcommand{\threadOverheadBatchOneDLRMBottomPZero}{4.6887}
\newcommand{\threadOverheadBatchOneDLRMBottomPOneHundred}{5.5817}
\newcommand{\globalOverheadBatchOneDLRMBottomPZero}{22.9803}
\newcommand{\globalOverheadBatchOneDLRMBottomPOneHundred}{24.0973}
\newcommand{\hybridOverheadBatchOneDLRMBottomPZero}{4.6314}
\newcommand{\hybridOverheadBatchOneDLRMBottomPOneHundred}{5.5109}
\newcommand{\threadOverheadBatchOneDLRMTopPZero}{5.4093}
\newcommand{\threadOverheadBatchOneDLRMTopPOneHundred}{5.9022}
\newcommand{\globalOverheadBatchOneDLRMTopPZero}{17.6910}
\newcommand{\globalOverheadBatchOneDLRMTopPOneHundred}{18.4404}
\newcommand{\hybridOverheadBatchOneDLRMTopPZero}{5.4621}
\newcommand{\hybridOverheadBatchOneDLRMTopPOneHundred}{5.9585}
\newcommand{\threadOverheadBatchTwentyFourtyEightDLRMBottomPZero}{8.2935}
\newcommand{\threadOverheadBatchTwentyFourtyEightDLRMBottomPOneHundred}{8.7873}
\newcommand{\globalOverheadBatchTwentyFourtyEightDLRMBottomPZero}{15.7861}
\newcommand{\globalOverheadBatchTwentyFourtyEightDLRMBottomPOneHundred}{16.5057}
\newcommand{\hybridOverheadBatchTwentyFourtyEightDLRMBottomPZero}{6.8432}
\newcommand{\hybridOverheadBatchTwentyFourtyEightDLRMBottomPOneHundred}{7.1807}
\newcommand{\threadOverheadBatchTwentyFourtyEightDLRMTopPZero}{10.5346}
\newcommand{\threadOverheadBatchTwentyFourtyEightDLRMTopPOneHundred}{10.9454}
\newcommand{\globalOverheadBatchTwentyFourtyEightDLRMTopPZero}{11.1176}
\newcommand{\globalOverheadBatchTwentyFourtyEightDLRMTopPOneHundred}{11.3891}
\newcommand{\hybridOverheadBatchTwentyFourtyEightDLRMTopPZero}{6.8507}
\newcommand{\hybridOverheadBatchTwentyFourtyEightDLRMTopPOneHundred}{7.0180}

\newcommand{\threadOverheadBatchSixtyFourNoscopeAmsterdamPZero}{4.0116}
\newcommand{\threadOverheadBatchSixtyFourNoscopeAmsterdamPOneHundred}{5.2361}
\newcommand{\globalOverheadBatchSixtyFourNoscopeAmsterdamPZero}{5.4722}
\newcommand{\globalOverheadBatchSixtyFourNoscopeAmsterdamPOneHundred}{9.9611}
\newcommand{\hybridOverheadBatchSixtyFourNoscopeAmsterdamPZero}{3.1402}
\newcommand{\hybridOverheadBatchSixtyFourNoscopeAmsterdamPOneHundred}{6.5644}
\newcommand{\threadOverheadBatchSixtyFourNoscopeCoralPZero}{4.1302}
\newcommand{\threadOverheadBatchSixtyFourNoscopeCoralPOneHundred}{5.3425}
\newcommand{\globalOverheadBatchSixtyFourNoscopeCoralPZero}{16.6478}
\newcommand{\globalOverheadBatchSixtyFourNoscopeCoralPOneHundred}{17.3212}
\newcommand{\hybridOverheadBatchSixtyFourNoscopeCoralPZero}{4.1185}
\newcommand{\hybridOverheadBatchSixtyFourNoscopeCoralPOneHundred}{5.3500}
\newcommand{\threadOverheadBatchSixtyFourNoscopeRoundaboutPZero}{1.6975}
\newcommand{\threadOverheadBatchSixtyFourNoscopeRoundaboutPOneHundred}{2.1009}
\newcommand{\globalOverheadBatchSixtyFourNoscopeRoundaboutPZero}{9.5159}
\newcommand{\globalOverheadBatchSixtyFourNoscopeRoundaboutPOneHundred}{9.7423}
\newcommand{\hybridOverheadBatchSixtyFourNoscopeRoundaboutPZero}{1.7021}
\newcommand{\hybridOverheadBatchSixtyFourNoscopeRoundaboutPOneHundred}{2.0999}
\newcommand{\threadOverheadBatchSixtyFourNoscopeTaipeiPZero}{2.9248}
\newcommand{\threadOverheadBatchSixtyFourNoscopeTaipeiPOneHundred}{3.9138}
\newcommand{\globalOverheadBatchSixtyFourNoscopeTaipeiPZero}{4.9376}
\newcommand{\globalOverheadBatchSixtyFourNoscopeTaipeiPOneHundred}{7.7799}
\newcommand{\hybridOverheadBatchSixtyFourNoscopeTaipeiPZero}{1.9543}
\newcommand{\hybridOverheadBatchSixtyFourNoscopeTaipeiPOneHundred}{4.0621}

%% file: abstract.tex
\begin{abstract}
\jack{Abstract is limited to 150 words for final version. Arxiv version in comments below. Full paper can be 12 pages}
Neural networks (NNs) are increasingly employed in safety-critical domains and in environments prone to unreliability (e.g., soft errors), such as on spacecraft. Therefore, it is critical to impart fault tolerance to NN inference. Algorithm-based fault tolerance (ABFT) is emerging as an efficient approach for fault tolerance in NNs.

We propose an adaptive approach to ABFT for NN inference that exploits untapped opportunities in emerging deployment scenarios. GPUs have high compute-to-memory-bandwidth ratios, while NN layers have a wide range of arithmetic intensities. This leaves some layers compute bound and others memory-bandwidth bound, but current approaches to ABFT do not consider these differences. We first investigate ABFT schemes best suited for each of these scenarios. We then propose \textit{intensity-guided ABFT}, an adaptive, arithmetic-intensity-guided approach that selects the most efficient ABFT scheme for each NN layer. Intensity-guided ABFT reduces execution-time overhead by 1.09--5.3$\times$ across many NNs compared to traditional approaches to ABFT.

\end{abstract}

%% file: intro.tex
\section{Introduction}\label{sec:intro}
Neural networks (\nns) are widely deployed for applications such as content recommendation~\cite{naumov2019deep}, medical diagnosis~\cite{candle-anl}, autonomous navigation~\cite{nvidia-drive}, and space imaging~\cite{denby2020orbital}. These applications desire \nns to reliably make correct predictions: mispredictions in content recommendation can lead to revenue loss~\cite{zhao2019aibox,li2021efficient}, while those in safety-critical applications can result in loss of life~\cite{iso-26262}.

One cause of unreliability in \nns is \ses: transient errors that occur in processing logic and memory in computing systems that can result in erroneous execution (e.g., $2 + 2 = 5$)~\cite{geist2016supercomputing}. The erroneous execution resulting from a \se is referred to as a fault. There are many causes of \ses, such as atmospheric radiation,  voltage scaling, hardware wearout, and manufacturing error~\cite{mittal2015survey,zamani2019greenmm,dixit2021silent}. Recent works have shown the potentially-catastrophic effects of \ses on \nns through fault injection~\cite{li2017understanding,chen2019binfi,chen2020tensorfi,mahmoud2020pytorchfi} and neutron beam experiments~\cite{dos2019reliability}: faults resulting from \ses can cause mispredictions in \nns at a rate that violates  automotive safety standards~\cite{li2017understanding,dos2019reliability,iso-26262}. Furthermore, the rate at which \ses occur increases with altitude~\cite{mittal2015survey,geist2016supercomputing} and in space~\cite{campbell1992single}, posing a challenge to the trend of deploying \nns on low-cost hardware on spacecraft~\cite{label2018nasa,denby2020orbital,hpwire-gpu-space}.

Therefore, applications that demand high reliability must employ some means of tolerating faults. However, tolerating faults caused by \ses requires performing redundant execution (\eg replication and comparison). For fault tolerance to be practical, it is critical that redundant execution operate with low overhead in terms of execution time and cost. 

In this work, we focus on software-based approaches for detecting faults that occur in processing logic during \nn inference on GPUs. We focus on detection, rather than correction, as detecting a catastrophic event is often more important to an application than quickly proceeding after such an event~\cite{hari2021making}. We focus on GPUs because they are commonly used for \nn inference in both cluster and edge settings, including in emerging space applications~\cite{denby2020orbital,hpwire-gpu-space}. We focus on faults that occur in processing logic, rather than in the memory hierarchy, as many modern systems contain ECC-protected memory hierarchies~\cite{nvidia-volta}. In contrast, processing logic is not as amenable to lightweight hardware fault tolerance~\cite{bartlett2004commercial}. %

\textit{Algorithm-based fault tolerance} (\abft\footnote{While we focus on detecting errors, an approach sometimes termed ``algorithm-based error detection'' (ABED), we use the more common terminology \abft for familiarity.}) is emerging as a promising approach toward imparting efficient software-based fault tolerance to \nn inference~\cite{ozen2019sanity,zhao2021ft,hari2021making,li2021efficient}. \abft adds redundant computations employing carefully-designed mathematical structures, and exploits the invariants so introduced to detect faults. This approach enables \abft to achieve significantly lower execution-time overhead than replication-based approaches. For this reason, \abft is a common approach for fault tolerance in traditional HPC computations, such as \matmatmult (\eg ~\cite{huang1984algorithm,bosilca2009algorithm,braun2014abft,zamani2019greenmm}), LU decomposition~\cite{wu2016towards}, sorting~\cite{li2019ft}, and other iterative methods~\cite{chen2013online}. 

\begin{figure}[t]
    \centering
    \includegraphics[width=0.9\linewidth]{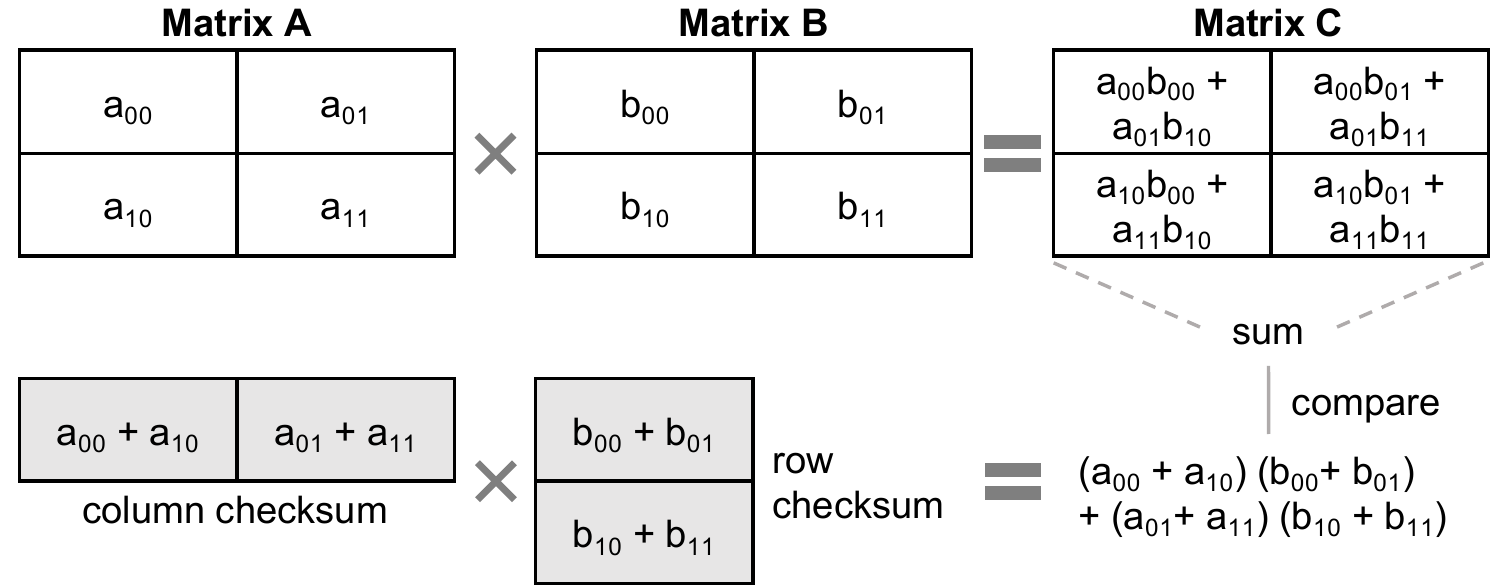}
    \vspace{-0.1in}
    \caption{Toy example of \abft with $\mathbf{\gemmM = \gemmN = \gemmK = 2}$.}
    \label{fig:abft-example}
    \vspace{-0.2in}
\end{figure}

\Figure\ref{fig:abft-example} shows a toy example of \abft-protected \matmatmult between matrices $\matA$ and $\matB$ of size $2\times2$ to produce output matrix $\matC$. \abft constructs a \textit{\checksumA vector} by summing each column of matrix $\matA$ and a \textit{\checksumB vector} by summing each row of matrix $\matB$. It is straightforward to see that the result of taking the dot product of these checksum vectors should, in the absence of a fault, equal the summation of all entries of output matrix $\matC$, which we refer to as the \checksumC. Correspondingly, comparison between the checksum dot-product result and the \checksumC can detect a single fault in $\matC$. 

Multiple recent works have explored leveraging \abft to impart fault tolerance to \nns~\cite{ozen2019sanity,zhao2021ft,hari2021making,li2021efficient}. Since existing \abft techniques support only linear computations, these approaches use \abft for  the linear operations of \nns (\eg fully-connected and convolutional layers, which are often executed as \matmatmults), and replicate nonlinear operations (\eg activation functions). We similarly focus on using \abft for linear layers implemented as \matmatmults in this work, and use the terminology ``\linearlayer'' to refer to fully-connected and convolutional layers.

Key to efficient operation in any approach to redundant execution is identifying and exploiting underutilized resources. If the computation-to-be-protected underutilizes certain compute units, redundant execution can potentially be performed on those units without adding much execution-time overhead. However, existing approaches to \abft typically only assume that computations being protected are compute bound, and thus aim to minimize the amount of redundant computation they perform. 

In this work, we first present a case for challenging this assumption based on trends in GPU hardware and \nn design: The introduction of processing units optimized for \nns (\eg \tensorcores~\cite{nvidia-tensorcores}) has led to an unprecedented increase in \FLOPS in inference-optimized GPUs. However, such GPUs have had a far less profound growth in memory bandwidth. This results in inference-optimized GPUs having high compute-to-memory-bandwidth ratios (\cmrs). High \cmrs require kernels to have high \ai to keep computational units highly utilized. However, many convolutional and fully-connected layers in \nns have \textit{low \ai}. Furthermore, many efforts toward reducing \nn latency, such as efficient \nn design~\cite{tan2019efficientnet}, model specialization~\cite{shen2017fast,kang2017noscope,hsieh2018focus,mullapudi2019online,kang2020blazeit}, and pruning~\cite{blalock2020state}, further reduce \ai.

These trends result in many \linearlayers in \nns that have \ai far lower than the \cmr on GPUs, rendering such layers memory-bandwidth\footnote{We refer to memory-bandwidth-bound layers as ``bandwidth-bound'' for short.} bound, rather than compute bound. As a result, such layers are unable to keep computational units highly utilized, opening opportunities for redundant execution to be performed for free. However, current approaches to \abft for \nn inference, which are well-suited for compute-bound \linearlayers, cannot exploit this opportunity to squeeze in redundant execution alongside the computation being protected.

To better exploit this nascent opportunity, we (1) investigate \abft schemes, which we refer to as \threadABFT, that exploit the unused computation cycles of the linear layer under protection on inference-optimized GPUs, and (2) propose a new, adaptive approach to \abft, called \hybridABFT, that selects among \threadABFT and traditional approaches to \abft on a per-\layer basis, using the \layer's \ai as a guide. %

To design an approach to \abft that can exploit the unused computation cycles of \linearlayers on modern inference-optimized GPUs, the key approach we leverage is to perform \abft at the smallest unit of the parallel subproblem performed by the \matmatmult for a \layer. As illustrated in \Figure\ref{fig:mm-decomposition}, high-performance \matmatmult on GPUs involves decomposing the overall \matmatmult into a hierarchy of subproblems across \threadblocks, \warps, and, at the smallest level, threads. Existing approaches to \abft for \nn inference on GPUs, which we term ``\globalABFT,'' generate checksums over the full input matrices to minimize the amount of redundant computation performed in checksum dot products. In contrast, we leverage an \abft scheme in which each thread performs \abft over the small \matmatmultShort subproblem it is responsible for. We refer to this approach as \textit{\threadABFT}. Under \threadABFT, each thread computes \abft checksums and dot products on the fly in tandem with its computation of the original \matmatmultShort, and performs its own thread-local checksum equality check.  

The approach taken in \threadABFT may at first appear counterintuitive, as it performs more redundant computation than \globalABFT: \threadABFT performs \abft over many small, thread-local \matmatmultsShort, whereas \globalABFT performs \abft over one large \matmatmultShort. In fact, \threadABFT results in multiple threads each computing identical checksums (\eg in \Figure\ref{fig:abft-example}, identical \checksumAs for threads that compute elements in the same rows in $\matC$). However, we show that, through careful design decisions, this approach is effective in exploiting the gaps in compute utilization of bandwidth-bound \linearlayers.  This approach also eliminates any additional loads/stores, which would compete with the \matmatmult itself for memory bandwidth, which is the bottleneck resource. The net result is low execution-time overhead for bandwidth-bound \linearlayers.

As described above, \threadABFT primarily benefits \linearlayers that are bandwidth bound. In contrast, it is not well-suited for compute-bound \linearlayers, for which \globalABFT suffices. As we show in \Section\ref{sec:opp}, \nns contain \textit{both bandwidth- and compute-bound \linearlayers}, making one-size-fits-all approaches inefficient.

Therefore, we propose \textit{\hybridABFT}, an adaptive \abft approach that selects among \globalABFT and \threadABFT for each \linearlayer of a \nn depending on which approach offers the lowest execution-time overhead, letting the \ai of the \layer and \cmr of the device guide such selection.

We implement and evaluate \hybridABFT atop CUTLASS~\cite{nvidia-cutlass}, a high-performance library from NVIDIA for \matmatmultsShort on GPUs. We evaluate  execution-time overhead on the inference-optimized NVIDIA T4 GPU when using \tensorcores. We consider eight popular convolutional \nns (\cnns), two \nns used within recommendation models (\dlrm)~\cite{naumov2019deep}, and four \cnns developed through model specialization and used for video analytics~\cite{kang2017noscope}. Compared to an optimized \globalABFT approach~\cite{hari2021making}, \hybridABFT reduces execution-time overhead by  up to 2.75$\times$ for popular \cnns, up to  4.55$\times$ for \dlrms, and up to 5.3$\times$ for specialized \cnns.  %
{These results show the promise of taking an arithmetic-intensity-guided approach to \abft to impart low-overhead fault tolerance to  \nn inference.}

\cutcandidate{
The contributions of this work are as follows:
\begin{denseitemize}
\item Presenting a case that key trends in GPU hardware and \nn design lead to many \linearlayers of \nns being memory-bandwidth-bound. This opens new, currently unexploited opportunities for efficient redundant execution. 

\item Investigating a thread-level approach to \abft for inference-optimized GPUs to exploit unused computational cycles in the bandwidth-bound \linearlayers of \nns.

\item Proposing \hybridABFT, which selects among \globalABFT and \threadABFT for each individual \linearlayer of a \nn, using the \ai of the \layer \cmr of the GPU as a guide.

\item Implementing and evaluating \hybridABFT across a wide range of \nn workloads, resulting in up to 5.3$\times$ reduction in execution-time overhead.
\end{denseitemize}
}

The code used in this paper is available at \url{https://github.com/Thesys-lab/arithmetic-intensity-guided-abft}.

%% file: background.tex
\section{Background} \label{sec:background}
In this section, we provide background on \matmatmultShort on GPUs, the need for fault tolerance in \nn inference, and how \abft is performed and optimized for \nn inference.%

\begin{figure*}[t]
    \centering
    \includegraphics[width=0.6\linewidth]{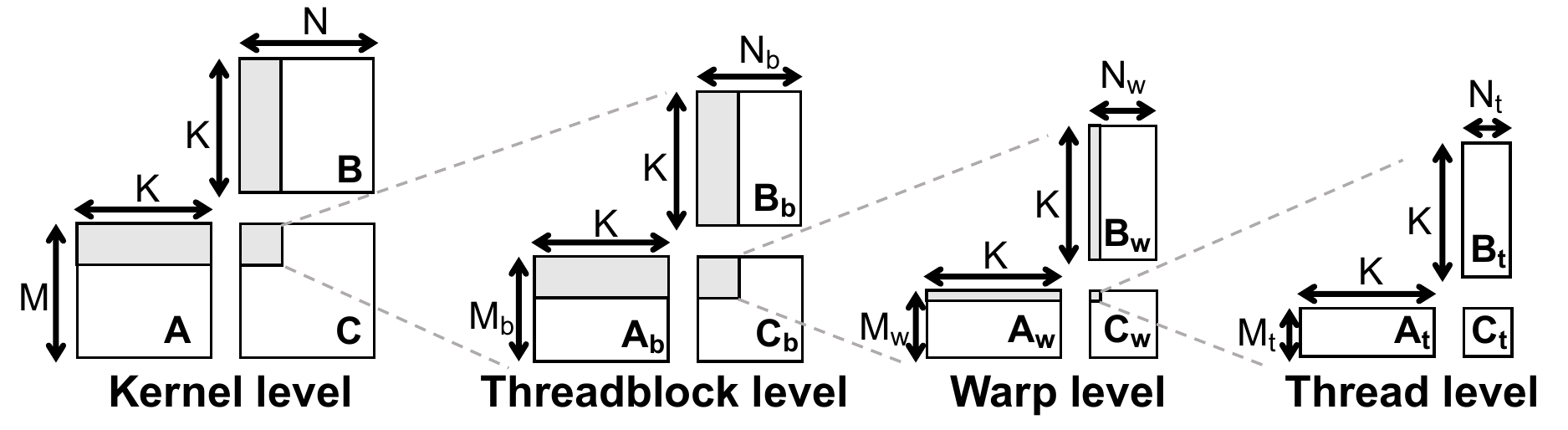}
    \caption{Hierarchical \matmatmultShort. Shaded regions show inputs/outputs used in the next level of the hierarchy.}
    \label{fig:mm-decomposition}
\end{figure*}

\subsection{Efficient matrix multiplication on GPUs} \label{sec:background:mm}
As described in \Section\ref{sec:intro}, our focus is on redundant execution for the convolutional and fully-connected layers of \nns, which we refer to as ``\linearlayers.'' For the remainder of this paper, we describe these operations as \matmatmults, as high-performance implementations of these \layers are often achieved through \matmatmults~\cite{nvidia-cutlass}. However, the approaches we propose can apply to other implementations as well. 

Within this setting, we denote a \linearlayer as the multiplication of matrix $\matA$ of size $\gemmM \times \gemmK$ by matrix $\matB$ of size $\gemmK \times \gemmN$ to produce an output matrix $\matC$ of size $\gemmM \times \gemmN$. Matrix $\matA$ contains the inputs to the \layer (\eg activations from the previous \layer). Matrix $\matB$ contains the learned weights of this \layer. Weights (matrix $\matB$) are known a priori, while activations (matrix $\matA$) are known only during computation. Output $\matC$ contains the output of the \layer, which will be fed to the next \layer, typically after being operated on by an activation function (\eg ReLU). 

\textbf{GPU terminology.} We use NVIDIA's terminology~\cite{nvidia-cuda} in describing the architectural components and programming abstractions of GPUs. A GPU consists of a number of \smsLong (\sms), each of which has many cores on which computation is performed along with a register file and shared memory region. Computation is executed on GPUs in \kernels consisting of many threads. Threads are grouped into \threadblocks, with all threads in a \threadblock executing on the same \sm and able to communicate with one another via shared memory. Groups of 32 threads within a \threadblock execute in lockstep as a so-called \warp. Each thread executes on an individual core, except when using \tensorcores, new processing units that enable \warp-wide collaborative execution of \matmatmults~\cite{nvidia-tensorcores}.

\textbf{Hierarchical matrix multiplication.}
High-performance implementations of \matmatmult on GPUs decompose the problem solved by the \kernel into a number of sub-matrix multiplications solved by \threadblocks, \warps, and threads. \Figure\ref{fig:mm-decomposition} shows an example of this decomposition: each \threadblock is responsible for computing a subset of $\matC$, which it decomposes into subsets to be computed by \warps of the \threadblock, each of which in turn decomposes the problem into subsets to be computed by individual threads. We denote the portions of $\matA$ and $\matB$ used by a thread as $\matAThread$ and $\matBThread$, respectively.
$\matAThread$ is of size $\gemmMThread \times \gemmK$ and $\matBThread$ is of size $\gemmK \times \gemmNThread$. 

\begin{figure}[t]
    \centering
    \includegraphics[width=0.35\linewidth]{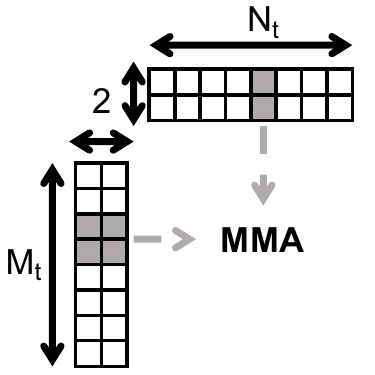}
    \caption{One step along the $\mathbf{\gemmK}$ dimension for a thread using  \mSixteenNeightKeight \tensorcore operations (\mmas). A total of \numThreadMMABold \mmas are performed per step, one for each  combination of two consecutive rows of $\mathbf{\matAThread}$ and one column of $\mathbf{\matBThread}$.%
    }
    \label{fig:m16n8k8}
\end{figure}

As our focus is on \nn inference, we focus on low-precision (\eg FP16) \matmatmults on \tensorcores, which are heavily used for accelerating inference. Our description follows the use of such operations in \cutlass. We focus in particular on the FP16 \mSixteenNeightKeight \tensorcore operation, though our discussion and proposed solutions apply to other \tensorcore operations as well.

Each \mSixteenNeightKeight \tensorcore operation is a \warp-wide operation that multiplies a $16\times8$ matrix $\matAtc$ by an $8\times8$ matrix $\matBtc$ and accumulates results into a $16\times8$ output matrix $\matCtc$ (we use subscript ``tc'' to denote \tensorcore operands/outputs)~\cite{nvidia-m16n8k8}. Each thread in the \warp provides four elements of $\matAtc$ and two elements of $\matBtc$ to the operation, and obtains four output elements of $\matCtc$ from the operation. 
We refer to one such \mSixteenNeightKeight matrix-multiply-accumulation operation as an ``\mma,'' following NVIDIA's terminology~\cite{nvidia-m16n8k8}.

\cutlass leverages \mmas within the hierarchical \matmatmultShort framework described above. Each thread walks down the $\gemmK$ dimension of the problem and loads an $\gemmMThread \times 2$ chunk of $\matAThread$ and a $2 \times \gemmNThread$ chunk of $\matBThread$. These loaded chunks are then used in \numThreadMMA \mmas, each of which uses two rows of the loaded chunk of $\matAThread$ and one column from the loaded chunk of $\matBThread$ from each thread, as shown in \Figure\ref{fig:m16n8k8}. The results of these operations are accumulated into the thread's \numThreadOutReg registers that store the partial accumulation of the thread's \matmatmultShort output. \cutlass uses standard optimizations to overlap loading the next chunks of $\matAThread$ and $\matBThread$ while the current \mmas are performed (\eg double buffering).

\revised{\subsection{Need for fault tolerance}
As described in \Section\ref{sec:intro}, our focus in this work is on detecting faults resulting from transient soft errors on GPUs. Handling soft errors has long been a concern for HPC systems~\cite{geist2016supercomputing,hochschild2021cores,dixit2021silent} due to their large scale and the criticality of the workloads that utilize them. %
Beyond these settings, the increasing trend of leveraging \nns in cyber-physical systems, such as autonomous vehicles~\cite{saxena2018keynote}, and in harsh operating environments, such as in spacecraft~\cite{hpe-gpu-space,hpwire-gpu-space,denby2020orbital}, has bolstered the need for fault tolerance solutions for \nns. For example, autonomous vehicles leverage GPUs and must tolerate faults to meet strict reliability requirements~\cite{iso-26262,saxena2018keynote}. Furthermore, servers equipped with general-purpose GPUs have recently been sent to the International Space Station to perform scientific computations~\cite{hpe-gpu-space,hpwire-gpu-space}, and which leverage software-based fault tolerance to handle the harsh operating environment therein.

Combining the longstanding need for reliability in HPC applications with the growing need for reliability in safety-critical and edge deployments of GPUs, efficient approaches to fault tolerance are necessary  both in the present and for the future.}

\subsection{Fault model} \label{sec:fault_model}
We next shift our focus to fault detection for \matmatmultShort. To set the stage, we first describe the fault model we consider.

We focus on detecting a single faulty output value in matrix $\matC$. We focus on detection, rather than correction, as being able to detect a catastrophic event is often more important than being able to quickly continue after such an event~\cite{hari2021making}. Following prior work~~\cite{ozen2019sanity,zhao2021ft,chen2021ranger}, we focus on detecting a single fault because the execution of one \layer in a \nn is short enough that the likelihood of more than one \se occurring during execution is low. %

We focus on faults occurring due to \ses in the processing logic of a GPU. We do not focus on faults in the memory hierarchy, such as in global memory, caches, shared memory, register files, and busses, as these components are more easily protected by ECC~\cite{nvidia-volta}. In contrast hardware fault tolerance for processing units is more expensive, typically requiring dual-modular-redundant circuits~\cite{bartlett2004commercial}. We also assume that control logic on the GPU is protected. This fault model is in line with prior work~\cite{li2018modeling,chen2021ranger}.
 
\subsection{\abft for matrix multiplication} \label{sec:background:abft}
\abft falls under a class of techniques called ``redundant execution'' in which additional computation is performed on top of the computation-to-be-protected for the purpose of fault tolerance. As described in \Section\ref{sec:intro}, \abft adds redundant computations employing carefully-designed mathematical structures, and exploits the invariants so introduced to detect errors while performing less redundant computation than replication-based approaches~\cite{huang1984algorithm}. 

\abft for \matmatmultShort typically operates by (1)~generating a $1 \times \gemmK$ \checksumA vector of matrix $\matA$ and a $\gemmK \times 1$ \checksumB vector of matrix $\matB$, (2)~performing the dot product between the \checksumA vector and \checksumB vector, (3)~summing all entries of the output matrix $\matC$, and (4)~comparing the values generated in (2) and (3) above. Approaches to \abft typically generate a single \checksumA for the entire input matrix $\matA$ (and similarly for $\matB$)~\cite{hari2021making,zhao2021ft}. We thus term such approaches ``\globalABFT.'' \GlobalABFT results in the minimum additional dot-product computations required for fault detection in \matmatmultShort, making it well-suited for compute-bound \matmatmultsShort. 

\revised{While we focus on detecting a single fault, \abft also supports detecting multiple faults. To do so, \abft generates multiple checksum columns and rows based on independent linear combinations of columns/rows. In this scenario, multiple output checksums are also generated based on these linear combinations and compared to checksum dot products. The approaches to \abft that we propose in this work can also handle higher fault rates in this way. %
}
 
\subsection{Optimizing \globalABFT for \nn inference} \label{sec:background:abft_nn}
Recent works leverage global \abft to protect the \linearlayers of \nns, and add multiple \nn-specific optimizations~\cite{hari2021making,zhao2021ft,li2021efficient}, which we describe next. Recall that, for \nn inference, matrix $\matA$ contains input activations and $\matB$ contains \layer weights. We therefore  refer to the \checksumA of matrix $\matA$ as the ``\actChecksum'' and the \checksumB of matrix $\matB$ as the ``\weightChecksum.'' 
 
\textit{Offline construction of \weightChecksum.} Since operand $\matB$ contains the \layer's weights, which remain the same for every inference request, the \weightChecksum of each \linearlayer in a \nn can be constructed once offline and reused for every inference request~\cite{hari2021making,zhao2021ft,li2021efficient}. The same does not hold for the \actChecksum of operand $\matA$, because its contents change for each inference request.
 
\textit{Checksum fusion.} Recent work~\cite{hari2021making} proposes to fuse the generation of the \checksumC used in the \abft check to the end of the \matmatmultShort \kernel itself. Kernel fusion significantly reduces the amount of data that must be read from memory to form the \checksumC, which speeds up checksum generation. As the next \layer's input $\matA$ is generated by the current \layer, the current \layer can also fuse the generation of the next layer's \actChecksum to the end of its \matmatmultShort kernel (after the activation function has been applied)~\cite{hari2021making}.

\textbf{Flow of \abft in \nn inference.} With the above optimizations, the workflow of an \abft-protected \linearlayer is as follows: (1)~perform \matmatmultShort to generate output $\matC$, (2)~perform fused \checksumC generation, (3)~apply the \layer's activation function to $\matC$, (4)~perform fused next-layer \actChecksum generation, (5)~launch a kernel that performs the \abft dot product for the current \layer and compares the results to the output checksum generated in Step~3. Steps 1--4 must take place sequentially, while Step~5 can take place in parallel with the next \layer of the \nn. Step~5 occurs in a separate kernel because it involves a global reduction over the partial checksums generated by \threadblocks.

By minimizing redundant computation, \globalABFT offers low execution-time overhead for compute-bound \linearlayers. However, we next identify trends in GPU hardware and \nns that lead to many \linearlayers being memory-bandwidth-bound. This opens new opportunities for efficient redundant execution that current approaches to \abft for \nn inference are unable to exploit. 

%% file: opportunity.tex
\section{New opportunities for efficient redundant execution} \label{sec:opp}
Critical to reducing execution-time overhead for any approach to redundant execution is discovering opportunities to exploit unused resources. In this section, we identify trends in GPU hardware and \nn design that create new, currently unexploited opportunities for efficient redundant execution in \nn inference. %

\subsection{Resource bottlenecks for GPU \kernels} \label{sec:opp:bottleneck}
GPU \kernels are typically either bound by computational throughput or by memory bandwidth. A popular model to determine whether a \kernel is compute or memory-bandwidth bound is comparing the the \textit{\ai} of the \kernel to the \textit{\cmrFull} (\cmr) of the device~\cite{williams2009roofline,nvidia-deep-learning-perf-guide}. Under this model, a \kernel is compute bound if the theoretical amount of time it spends performing computation is greater than the theoretical amount of time it spends loading/storing data from/to memory: 
\begin{align*}
    \frac{\text{\ops}}{\text{Compute Bandwidth}} & >  \frac{\text{Bytes}}{\text{Memory Bandwidth}}
\end{align*}
Here, ``\ops'' is the number of arithmetic operations performed by the \kernel, ``Bytes'' is the amount of data it transfers to/from memory, ``Compute Bandwidth'' is the GPU's peak \FLOPS, and ``Memory Bandwidth'' is the GPU's memory bandwidth (bytes/sec). Rearranging this inequality to pair properties of the \kernel on the left-hand side and properties of the GPU on the right-hand gives: 
\begin{align}
     \frac{\text{\ops}}{\text{Bytes}} & >  \frac{\text{Compute Bandwidth}}{\text{Memory Bandwidth}}
    \label{equation:arithmetic_intensity}
\end{align}
The left-hand ratio of Equation~\ref{equation:arithmetic_intensity} is the \kernel's \ai: the ratio between the \ops the \kernel performs and the bytes it transfers to/from memory. The right-hand ratio is the GPU's \cmr.

\textbf{Takeaway.} From the lens of this performance model, it is clear that the \aiShort of a given \kernel and \cmr of a given GPU play key roles in determining opportunities for redundant execution to leverage unused resources. For example, a \kernel with low \aiShort running on a GPU with a high \cmr will likely be bandwidth bound and underutilize compute units. This leaves opportunities for redundant execution to leverage such units without hampering the performance of the \kernel itself. 

We next examine trends in GPU hardware and \nn design to identify opportunities for such efficient redundant execution.

\subsection{Wide range of \aisShort exhibited by \nn \layers} \label{sec:opp:ai} 
We first examine the \aisShort of current \nns and their individual \linearlayers under various operational settings.
In this analysis, we consider only ``linear layers'', such as convolutional and fully-connected layers, which are often implemented as \matmatmults. Other operations, such as activation functions, are typically fused to these \linearlayers and contribute far less to overall \ai and execution time. 

The ``\aggregateAI'' of a \nn as a whole is computed by summing the \ops performed across all \linearlayers, summing the bytes read/written across all \linearlayers, and dividing these quantities. This metric provides an estimate of whether the \nn as a whole is more compute or memory-bandwidth bound.

\Figure\ref{fig:background:ai_normal} shows the FP16 \aggregateAIs of eight widely-used \cnns from the popular PyTorch Torchvision library~\cite{pytorch-torchvision-cnns}.\footnote{\updated{We replace the group convolutions in \nameShufflenet and \nameResnext with non-grouped convolutions to ease their conversion to \matmatmultsShort. The reported \aggregateAIs of these \nns are, thus, higher than they would be with grouped convolutions, which typically decrease \ai.}} The figure shows a wide range of \aggregateAIs among such \cnns (from 71 to 220). 

Furthermore, many domains leverage \nns that are significantly smaller than those described above, and thus have even lower \aggregateAIs. For example, \nns used for recommendation serving, such as Facebook's popular \dlrm~\cite{naumov2019deep}, leverage small \nns consisting of a few fully-connected \layers. Consequently, these \nns have low \aggregateAIs (\eg 7 in FP16). %

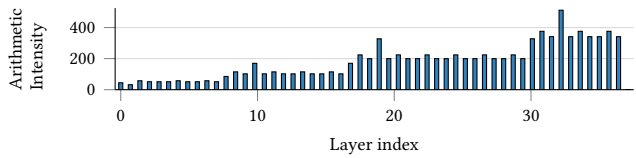
\begin{figure}[t]
    \centering
    \newcommand{\figureheight}{1.05in}
    \newcommand{\figurewidth}{\linewidth}
    \input{figures/background_conv_resnet50_ai}
    \vspace{-0.3in}
    \caption{FP16 \aiShort of convolutional and fully-connected \layers of \nameResnet on HD images (resolution $\mathbf{1080 \times 1920}$) with batch size of one.}
    \label{fig:background:ai_resnext}
    \vspace{-0.2in}
\end{figure}

\Figure\ref{fig:background:ai_resnext} shows the \aisShort of individual convolutional and fully-connected \layers of \nameResnet. As illustrated, there is a wide variance of \aisShort (1--511) among even various \linearlayers of the same \nn (other \nns are similar).

Finally, \aiShort also varies with settings of the applications in which \nns operate, such as the size of inputs to the \nn. For example, increasing the batch size used in inference typically increases \aiShort by amortizing the overhead of loading \nn weights from memory. Thus, the many applications that use small batch sizes for low-latency inference are likely to have low \aiShort~\cite{zhang2018deepcpu,chung2018serving}, while those that can aggressively batch inputs may have higher \aiShort. For example, the FP16 \aggregateAIs of the \nns used in \dlrm increase from 7 at batch size of 1 to 70--109 at batch size 256. For \cnns, the resolution of input images also affects \aiShort for similar reasons, as operating over large images amortizes the cost of loading convolutional filters from memory. For example, the FP16 \aggregateAI of \nameResnet is 72 when operating over images of resolution $224\times224$ (the resolution typically used for ImageNet~\cite{russakovsky2015imagenet}), but increases to 122 when operating over images of resolution $1080\times1920$ (typically considered HD). 

\textbf{Takeaway.} \Nns exhibit wide variance in \aiShort across \nns, across individual \linearlayers within a \nn, and across application settings. This renders some \nns, some layers, and some application settings likely to underutilize computational resources. 

\begin{figure}[t]
    \centering
    \newcommand{\figureheight}{1.05in}
    \newcommand{\figurewidth}{\linewidth}
    \input{figures/background_conv_normal_ai}
    \vspace{-0.3in}
    \caption{FP16 \aggregateAI of \cnns operating on images of size $\mathbf{1080 \times 1920}$ at batch size of one. %
    }
    \label{fig:background:ai_normal}
\end{figure}

\subsection{Inference-optimized GPUs have high \cmr} \label{sec:opp:cmr}
We now discuss trends in \cmr, the right-hand ratio in Equation~\ref{equation:arithmetic_intensity}.

General-purpose GPUs have been a workhorse for \nns since the early 2010s~\cite{krizhevsky2012imagenet}. Recent GPUs have further bolstered \nn acceleration by adding hardware units specifically designed for the \matmatmultsShort found in \nns, such as NVIDIA's \tensorcores~\cite{nvidia-tensorcores}. These hardware units offer unprecedented performance in terms of \FLOPS, particularly when using low-precision arithmetic (\eg FP16), as is common in \nn inference.

For example, the inference-optimized T4 GPU offers 65 FP16 \TFLOPS~\cite{nvidia-turing}, a marked increase from the 11 FP16 \TFLOPS offered by its predecessor, the P4~\cite{nvidia-p4}, which did not contain \tensorcores. Such high performance is also offered in other server-grade GPUs, such as the V100 and A100 GPUs, which offer 125 and 312 FP16 \TFLOPS, respectively~\cite{nvidia-volta,nvidia-ampere}. This trend has also made its way to edge devices, as GPUs in the NVIDIA Jetson family now offer up to 32 INT8 \TOPS via \tensorcores, whereas predecessors were bound to single-digit INT8 \TOPS~\cite{nvidia-jetson}.

While the \FLOPS offered by inference-optimized GPUs has drastically increased, memory bandwidth has not increased at the same rate. For example, while the T4 GPU increases FP16 \FLOPS by 5.9$\times$ compared to the P4 GPU, it offers only a 1.7$\times$ increase in memory bandwidth. Similar trends hold for other GPUs. 

The net result of these trends in compute and memory bandwidth is a \textit{significant increase in the \cmr of GPUs}. For example, the FP16 \cmr of the T4 GPU is \tFourHalfCMR, while that of the P4 was \pFourHalfCMR. Even GPUs with high-bandwidth memory (\eg HBM2) have high \cmrs (139 and 201 in FP16 for V100 and A100, respectively~\cite{nvidia-volta,nvidia-ampere}), as do edge GPUs (235 in INT8 for Jetson AGX Xavier~\cite{nvidia-jetson}).

\textbf{Takeaway.} The introduction of specialized hardware units for \matmatmultsShort that drastically increase computational throughput, but a significantly slower increase in memory bandwidth, results in inference-optimized GPUs with high \cmrs. This trend ``raises the bar'' for GPU \kernels, \textit{making them more likely to be memory-bandwidth bound and underutilize GPU compute units.}

\subsection{Many \nn optimization trends reduce \aiShort} \label{sec:opp:small_nn}
A secondary trend further exacerbates the growing bandwidth-bound nature of many \nns: designing small \nns to perform tasks with high throughput or low latency. 
The \nns shown in \Figure\ref{fig:background:ai_normal} are large, general-purpose \nns designed to classify a wide variety of objects (\eg from ImageNet~\cite{russakovsky2015imagenet}). There is a growing body of work on designing more efficient \nn architectures that can accomplish the same task as such general-purpose \nns, but with a significantly smaller \nn. There are many techniques along these lines, including efficient neural architecture search~\cite{zoph2016neural,tan2019efficientnet}, pruning~\cite{blalock2020state}, and model specialization~\cite{shen2017fast,kang2017noscope,hsieh2018focus,mullapudi2019online,kang2020blazeit}. These techniques often result in deploying \nns with lower \aggregateAI than the general-purpose \nns shown in \Figure\ref{fig:background:ai_normal}. 

For example, in model specialization for offline video analytics, a small, specialized \cnn is designed to answers specific queries (\eg find red trucks), and which consults a larger, general-purpose \cnn only when  unsure~\cite{shen2017fast,kang2017noscope,hsieh2018focus}. By targeting a focused query, specialized \cnns can typically be made smaller and faster than general-purpose \nns, but exhibit far lower \aggregateAI: the specialized \cnns from the widely-cited NoScope video analytics system~\cite{kang2017noscope} have FP16 \aggregateAIs of 15--53, even with large batch size.

\textbf{Takeaway.} Current trends in efficient \nn design result in \nns that have lower \aiShort, making current and future workloads likely to underutilize GPU compute units.

\subsection{Takeaways and new opportunities}
The previous sections have identified trends that lead to the conclusion that \textit{the current and future landscape of \nn inference will contain a significant number of memory-bandwidth bound \linearlayers}: \Section\ref{sec:opp:ai} illustrated that current \nns, the \linearlayers within them, and their application settings exhibit a wide variance of \aisShort (including many with low \aiShort), and \Section\ref{sec:opp:small_nn} described increasingly prevalent trends in \nn design that often reduce \aiShort. Coupling this with the dramatic growth in \cmr for GPUs described in \Section\ref{sec:opp:cmr} drives home the conclusion that current and future \nns will contain bandwidth-bound \linearlayers that underutilize the computational capabilities of GPUs. 

Such bandwidth-bound \linearlayers leave room open for redundant execution to fill gaps in compute utilization during \matmatmultShort. However, current approaches to \abft for \nn inference are unable to exploit these fine-grained opportunities for efficient redundant execution. As described in \Section\ref{sec:background:abft_nn}, \globalABFT operates at a much higher level (specifically, kernel level), and hence is unable to exploit compute underutilization that occurs at \textit{finer granularity within the \matmatmultShort operation.} 

This calls for investigating approaches to redundant execution that can exploit the fine-grained compute underutilization exhibited by current and future \matmatmultShort \kernels in \nn inference. Such an approach would complement \globalABFT, which is well-suited for the compute-bound \linearlayers in \nns. 

We next turn our focus toward investigating such an approach.

\textbf{Key design principle.} \label{sec:principle}
Driven by the opportunities outlined above, we use the following principle when considering approaches to redundant execution for memory-bandwidth-bound \matmatmultsShort: \textit{avoid performing additional memory accesses whenever possible even if doing so comes at the expense of additional computation}. Adhering to this principle avoids competing with the \matmatmultShort for its bottleneck resource, memory bandwidth.

%% file: figures/background_conv_normal_ai.tex
\begin{tikzpicture}[]
\pgfplotsset{label style={font=\footnotesize}, 
             tick label style={font=\footnotesize},
             legend style={font=\footnotesize},
             every non boxed x axis/.append style={x axis line style=-},
             every non boxed y axis/.append style={y axis line style=-},
             axis lines=left,
             /pgfplots/ybar legend/.style={
                /pgfplots/legend image code/.code={%
                    \draw[##1,/tikz/.cd,yshift=-0.25em]
                    (0cm,0cm) rectangle (12pt,6pt);
                },
            },
        }

\begin{axis}[
height=\figureheight,
tick align=outside,
tick label style={/pgf/number format/assume math mode},
tick pos=left,
width=\figurewidth,
x grid style={white!69.01960784313725!black},
xmin=-0.41, xmax=5.16,
xticklabel style={rotate=18, align=right},
xtick={0,0.7,1.4,2.1,2.8,3.5,4.2,4.9},
xticklabels={
{\nameSqueezenet},
{\nameShufflenet},
{\nameDensenet},
{\nameResnet},
{\nameAlexnet},
{\nameVgg},
{\nameResnext},
{\nameWideResnet}},
xlabel style={align=center,at={(axis description cs:0.5,-.2)}},
xlabel={Model},
y grid style={gray, opacity=0.3},
ylabel style={at={(axis description cs:0.01,.4)}},
ymajorgrids,
ylabel style={align=left},
ylabel={Aggregate arithmetic\\intensity},
ymin=0, ymax=250
]

\draw[draw=black,fill=color_blue_3_2] (axis cs:-0.15,0) rectangle (axis cs:0.15,\aiBatchOneResBigSqueezenet);
\draw[draw=black,fill=color_blue_3_2] (axis cs:0.55,0) rectangle (axis cs:0.85,\aiBatchOneResBigShufflenet);
\draw[draw=black,fill=color_blue_3_2] (axis cs:1.25,0) rectangle (axis cs:1.55,\aiBatchOneResBigDensenet);
\draw[draw=black,fill=color_blue_3_2] (axis cs:1.95,0) rectangle (axis cs:2.25,\aiBatchOneResBigResnet);
\draw[draw=black,fill=color_blue_3_2] (axis cs:2.65,0) rectangle (axis cs:2.95,\aiBatchOneResBigAlexnet);
\draw[draw=black,fill=color_blue_3_2] (axis cs:3.35,0) rectangle (axis cs:3.65,\aiBatchOneResBigVgg);
\draw[draw=black,fill=color_blue_3_2] (axis cs:4.05,0) rectangle (axis cs:4.35,\aiBatchOneResBigResnext);
\draw[draw=black,fill=color_blue_3_2] (axis cs:4.75,0) rectangle (axis cs:5.05,\aiBatchOneResBigWideResnet);

\end{axis}

\end{tikzpicture}

%% file: figures/background_conv_resnet50_ai.tex
\begin{tikzpicture}[]
\pgfplotsset{label style={font=\footnotesize}, 
             tick label style={font=\footnotesize},
             legend style={font=\footnotesize},
             every non boxed x axis/.append style={x axis line style=-},
             every non boxed y axis/.append style={y axis line style=-},
             axis lines=left,
             /pgfplots/ybar legend/.style={
                /pgfplots/legend image code/.code={%
                    \draw[##1,/tikz/.cd,yshift=-0.25em]
                    (0cm,0cm) rectangle (12pt,6pt);
                },
            },
        }

\begin{axis}[
height=\figureheight,
tick align=outside,
tick label style={/pgf/number format/assume math mode},
tick pos=left,
width=\figurewidth,
x grid style={white!69.01960784313725!black},
xmin=-0.41, xmax=37.51,
xlabel={Layer index},
y grid style={gray, opacity=0.3},
ylabel style={at={(axis description cs:0.01,.5)}},
ymajorgrids,
ylabel style={align=left},
ylabel={Arithmetic\\Intensity},
ymin=0, ymax=525
]

\draw[draw=black,fill=color_blue_3_2] (axis cs:-0.1500,0) rectangle (axis cs:0.1500 ,44.58384306071621 );
\draw[draw=black,fill=color_blue_3_2] (axis cs:0.5500,0) rectangle (axis cs:0.8500  ,31.992100715872624);
\draw[draw=black,fill=color_blue_3_2] (axis cs:1.2500,0) rectangle (axis cs:1.5500  ,57.57441137272323 );
\draw[draw=black,fill=color_blue_3_2] (axis cs:1.9500,0) rectangle (axis cs:2.2500  ,51.17978082732748 );
\draw[draw=black,fill=color_blue_3_2] (axis cs:2.6500,0) rectangle (axis cs:2.9500  ,51.17978082732748 );
\draw[draw=black,fill=color_blue_3_2] (axis cs:3.3500,0) rectangle (axis cs:3.6500  ,51.17978082732748 );
\draw[draw=black,fill=color_blue_3_2] (axis cs:4.0500,0) rectangle (axis cs:4.3500  ,57.57441137272323 );
\draw[draw=black,fill=color_blue_3_2] (axis cs:4.7500,0) rectangle (axis cs:5.0500  ,51.17978082732748 );
\draw[draw=black,fill=color_blue_3_2] (axis cs:5.4500,0) rectangle (axis cs:5.7500  ,51.17978082732748 );
\draw[draw=black,fill=color_blue_3_2] (axis cs:6.1500,0) rectangle (axis cs:6.4500  ,57.57441137272323 );
\draw[draw=black,fill=color_blue_3_2] (axis cs:6.8500,0) rectangle (axis cs:7.1500  ,51.17978082732748 );
\draw[draw=black,fill=color_blue_3_2] (axis cs:7.5500,0) rectangle (axis cs:7.8500  ,85.27718374732686 );
\draw[draw=black,fill=color_blue_3_2] (axis cs:8.2500,0) rectangle (axis cs:8.5500  ,114.79185119574845);
\draw[draw=black,fill=color_blue_3_2] (axis cs:8.9500,0) rectangle (axis cs:9.2500  ,102.07738505464212);
\draw[draw=black,fill=color_blue_3_2] (axis cs:9.6500,0) rectangle (axis cs:9.9500  ,169.77239233666285);
\draw[draw=black,fill=color_blue_3_2] (axis cs:10.3500,0) rectangle (axis cs:10.6500,102.07738505464212);
\draw[draw=black,fill=color_blue_3_2] (axis cs:11.0500,0) rectangle (axis cs:11.3500,114.79185119574845);
\draw[draw=black,fill=color_blue_3_2] (axis cs:11.7500,0) rectangle (axis cs:12.0500,102.07738505464212);
\draw[draw=black,fill=color_blue_3_2] (axis cs:12.4500,0) rectangle (axis cs:12.7500,102.07738505464212);
\draw[draw=black,fill=color_blue_3_2] (axis cs:13.1500,0) rectangle (axis cs:13.4500,114.79185119574845);
\draw[draw=black,fill=color_blue_3_2] (axis cs:13.8500,0) rectangle (axis cs:14.1500,102.07738505464212);
\draw[draw=black,fill=color_blue_3_2] (axis cs:14.5500,0) rectangle (axis cs:14.8500,102.07738505464212);
\draw[draw=black,fill=color_blue_3_2] (axis cs:15.2500,0) rectangle (axis cs:15.5500,114.79185119574845);
\draw[draw=black,fill=color_blue_3_2] (axis cs:15.9500,0) rectangle (axis cs:16.2500,102.07738505464212);
\draw[draw=black,fill=color_blue_3_2] (axis cs:16.6500,0) rectangle (axis cs:16.9500,169.77239233666285);
\draw[draw=black,fill=color_blue_3_2] (axis cs:17.3500,0) rectangle (axis cs:17.6500,224.07322654462243);
\draw[draw=black,fill=color_blue_3_2] (axis cs:18.0500,0) rectangle (axis cs:18.3500,199.7857689364958 );
\draw[draw=black,fill=color_blue_3_2] (axis cs:18.7500,0) rectangle (axis cs:19.0500,327.62860727728986);
\draw[draw=black,fill=color_blue_3_2] (axis cs:19.4500,0) rectangle (axis cs:19.7500,199.7857689364958 );
\draw[draw=black,fill=color_blue_3_2] (axis cs:20.1500,0) rectangle (axis cs:20.4500,224.07322654462243);
\draw[draw=black,fill=color_blue_3_2] (axis cs:20.8500,0) rectangle (axis cs:21.1500,199.7857689364958 );
\draw[draw=black,fill=color_blue_3_2] (axis cs:21.5500,0) rectangle (axis cs:21.8500,199.7857689364958 );
\draw[draw=black,fill=color_blue_3_2] (axis cs:22.2500,0) rectangle (axis cs:22.5500,224.07322654462243);
\draw[draw=black,fill=color_blue_3_2] (axis cs:22.9500,0) rectangle (axis cs:23.2500,199.7857689364958 );
\draw[draw=black,fill=color_blue_3_2] (axis cs:23.6500,0) rectangle (axis cs:23.9500,199.7857689364958 );
\draw[draw=black,fill=color_blue_3_2] (axis cs:24.3500,0) rectangle (axis cs:24.6500,224.07322654462243);
\draw[draw=black,fill=color_blue_3_2] (axis cs:25.0500,0) rectangle (axis cs:25.3500,199.7857689364958 );
\draw[draw=black,fill=color_blue_3_2] (axis cs:25.7500,0) rectangle (axis cs:26.0500,199.7857689364958 );
\draw[draw=black,fill=color_blue_3_2] (axis cs:26.4500,0) rectangle (axis cs:26.7500,224.07322654462243);
\draw[draw=black,fill=color_blue_3_2] (axis cs:27.1500,0) rectangle (axis cs:27.4500,199.7857689364958 );
\draw[draw=black,fill=color_blue_3_2] (axis cs:27.8500,0) rectangle (axis cs:28.1500,199.7857689364958 );
\draw[draw=black,fill=color_blue_3_2] (axis cs:28.5500,0) rectangle (axis cs:28.8500,224.07322654462243);
\draw[draw=black,fill=color_blue_3_2] (axis cs:29.2500,0) rectangle (axis cs:29.5500,199.7857689364958 );
\draw[draw=black,fill=color_blue_3_2] (axis cs:29.9500,0) rectangle (axis cs:30.2500,327.62860727728986);
\draw[draw=black,fill=color_blue_3_2] (axis cs:30.6500,0) rectangle (axis cs:30.9500,375.89251439539345);
\draw[draw=black,fill=color_blue_3_2] (axis cs:31.3500,0) rectangle (axis cs:31.6500,341.1103853690398 );
\draw[draw=black,fill=color_blue_3_2] (axis cs:32.0500,0) rectangle (axis cs:32.3500,511.4985308521058 );
\draw[draw=black,fill=color_blue_3_2] (axis cs:32.7500,0) rectangle (axis cs:33.0500,341.1103853690398 );
\draw[draw=black,fill=color_blue_3_2] (axis cs:33.4500,0) rectangle (axis cs:33.7500,375.89251439539345);
\draw[draw=black,fill=color_blue_3_2] (axis cs:34.1500,0) rectangle (axis cs:34.4500,341.1103853690398 );
\draw[draw=black,fill=color_blue_3_2] (axis cs:34.8500,0) rectangle (axis cs:35.1500,341.1103853690398 );
\draw[draw=black,fill=color_blue_3_2] (axis cs:35.5500,0) rectangle (axis cs:35.8500,375.89251439539345);
\draw[draw=black,fill=color_blue_3_2] (axis cs:36.2500,0) rectangle (axis cs:36.5500,341.1103853690398 );
\draw[draw=black,fill=color_blue_3_2] (axis cs:36.9500,0) rectangle (axis cs:37.2500,0.9985139304394631);

\end{axis}

\end{tikzpicture}

%% file: technique.tex
\section{Thread-level replication?} \label{sec:thread:replication}
A natural question that arises when considering options for redundant execution for bandwidth-bound \linearlayers is whether it is beneficial to use \textit{thread-level replication}, rather than \abft.  After all, \abft is primarily designed to reduce the number of redundant operations performed compared to replication, while spare compute cycles are plentiful in bandwidth-bound \linearlayers. Furthermore, thread-level replication easily satisfies the design principle stated in \Section\ref{sec:principle}, by sharing loads with the original \matmatmult.

We began our exploration of redundant execution for bandwidth-bound \linearlayers with replication for these very reasons, but ultimately found it to have higher execution-time overhead than \abft, as we next describe. We focus on matrix multiplications using \mSixteenNeightKeight FP16 \tensorcore operations (\mmas) (described in \Section\ref{sec:background:mm}). Recall that, in \matmatmultShort using this operation, each thread participates in \numThreadMMA \mmas on each iteration along the $\gemmK$ dimension. For each \mma, a thread provides four elements from $\matAThread$, two elements from $\matBThread$, and receives four output elements.

We have considered two approaches to thread-level replication:

\textbf{Traditional replication.} The traditional approach to performing thread-level replication is to perform \numThreadMMA additional \mmas per step down the $\gemmK$ dimension, accumulate the results in a separate set of \numThreadOutReg registers, and compare these registers to the original \numThreadOutReg \matmatmultShort output registers. However, we found that the $2\times$ increase in output register usage per thread in this approach limits the number of \threadblocks that can be co-scheduled on a single \sm (so-called ``occupancy''~\cite{luitjens2011occupancy}), and leads to significant slowdowns compared to the original \matmatmultShort \kernel. 

\textbf{Replicated \mma, single accumulation.} Based on this limitation, we next explored replicating \mmas, but accumulating results to a single set of four output registers. Under this approach, one still performs \numThreadMMA additional \mmas per step along the $\gemmK$ dimension, but each redundant \mma accumulates results to the same set of four registers. By the end of the thread-level \matmatmultShort, in the absence of a fault, the summation of these four registers will equal the summation of the thread's ``original'' \numThreadOutReg output registers. Threads can use this invariant to detect faults. 

We find that the limited additional register usage of this approach alleviates the occupancy-related slowdowns described above, and thus significantly reduces execution-time overhead compared to the traditional form of replication. However, as we will show in \Section\ref{sec:eval:square}, doubling the number of \mmas performed results in higher execution-time overhead than \abft. 

We thus turn our focus to investigating \abft schemes that can exploit the compute underutilization identified in \Section\ref{sec:opp}. 

\section{{Arithmetic-intensity-guided \abft}} \label{sec:thread}
In this section, we first investigate approaches to \abft that can exploit the fine-grained compute underutilization of bandwidth-bound \linearlayers identified in \Section\ref{sec:opp}. We then describe the design of an adaptive approach to \abft that selects an \abft scheme for each \linearlayer guided by the \layer's \aiShort. %

\subsection{At which level should \abft be performed?}
The hierarchical decomposition of \matmatmultsShort described in \Section\ref{sec:background:mm} offers multiple levels at which \abft can be performed: the \kernel level (as in \globalABFT), \threadblock level, \warp level, or thread level. However, performing \abft at any level other than the thread level requires performing additional loads/stores to generate checksums. For example, performing \abft at the level of a \threadblock requires individual threads to cooperate to generate \threadblock-wide checksums, which requires storing and loading thread-local partial checksums. Such additional loads and stores violate the design principle described in \Section\ref{sec:principle} and compete for bandwidth with the \matmatmultShort itself. %

\begin{figure}[t]
    \centering
    \includegraphics[width=0.9\linewidth]{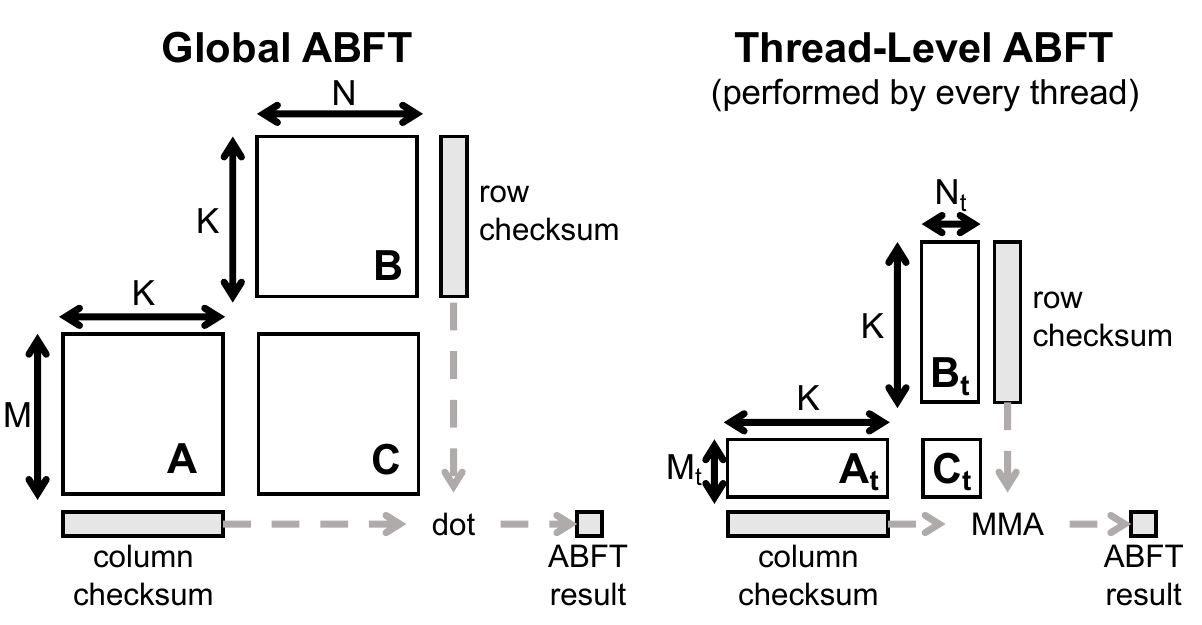}
    \caption{Global and two-sided thread-level \abft.}
    \label{fig:global_v_thread}
\end{figure}

In contrast, performing \abft at the level of individual threads avoids additional loads/stores. \Figure\ref{fig:global_v_thread} compares one approach to \threadABFT with \globalABFT at a high level. Concretely, \threadABFT involves threads in the \matmatmultShort \kernel performing their own, local \abft calculations across their own, local sub-\matmatmultsShort. \ThreadABFT eliminates additional loads/stores by (1)~sharing the loads of operands that will be used for checksum generation with those that were already performed for thread-level \matmatmultShort in a step along the $\gemmK$ dimension, and (2)~eliminating stores of partial checksums for use in \threadblock- or \warp-wide checksum generation. 

Thus, we conclude that performing \abft at the thread level is the appropriate fit for \abft optimized for bandwidth-bound \linearlayers. This conclusion is heavily driven by the design principle established in \Section\ref{sec:principle} of avoiding additional loads/stores. In cases where this principle can be relaxed, performing \abft at other levels of the \matmatmultShort hierarchy may be appropriate. Even with the somewhat-extreme stance taken in \threadABFT, we will show in \Section\ref{sec:eval}  that \threadABFT significantly reduces execution-time overhead for bandwidth-bound \linearlayers.

\subsection{Design decisions for \threadABFT} \label{sec:thread:design}
Even having narrowed our focus to performing \abft at thread level for bandwidth-bound \linearlayers, there remain multiple design decisions that affect performance, which we discuss next. Similar to  \Section\ref{sec:thread:replication}, we focus on \mSixteenNeightKeight \tensorcore operations (\mmas), which are described in detail in \Section\ref{sec:background:mm}.

\subsubsection{Online computation of \weightChecksums} 
Recall from \Section\ref{sec:background:abft_nn} that optimized approaches to \globalABFT for \nns typically compute the \weightChecksum of $\matB$ once offline and load it upon every inference request. We do not employ this technique for \threadABFT, as doing so would require threads to load \weightChecksums from memory, violating the design principle described in \Section\ref{sec:principle}. Thus, \threadABFT recomputes thread-local \weightChecksums alongside the thread-level \matmatmultShort.

\subsubsection{Balancing checksum generation and redundant \mmas} \label{sec:thread:checksum} 
\begin{figure}[t]
    \centering
    \includegraphics[width=0.9\linewidth]{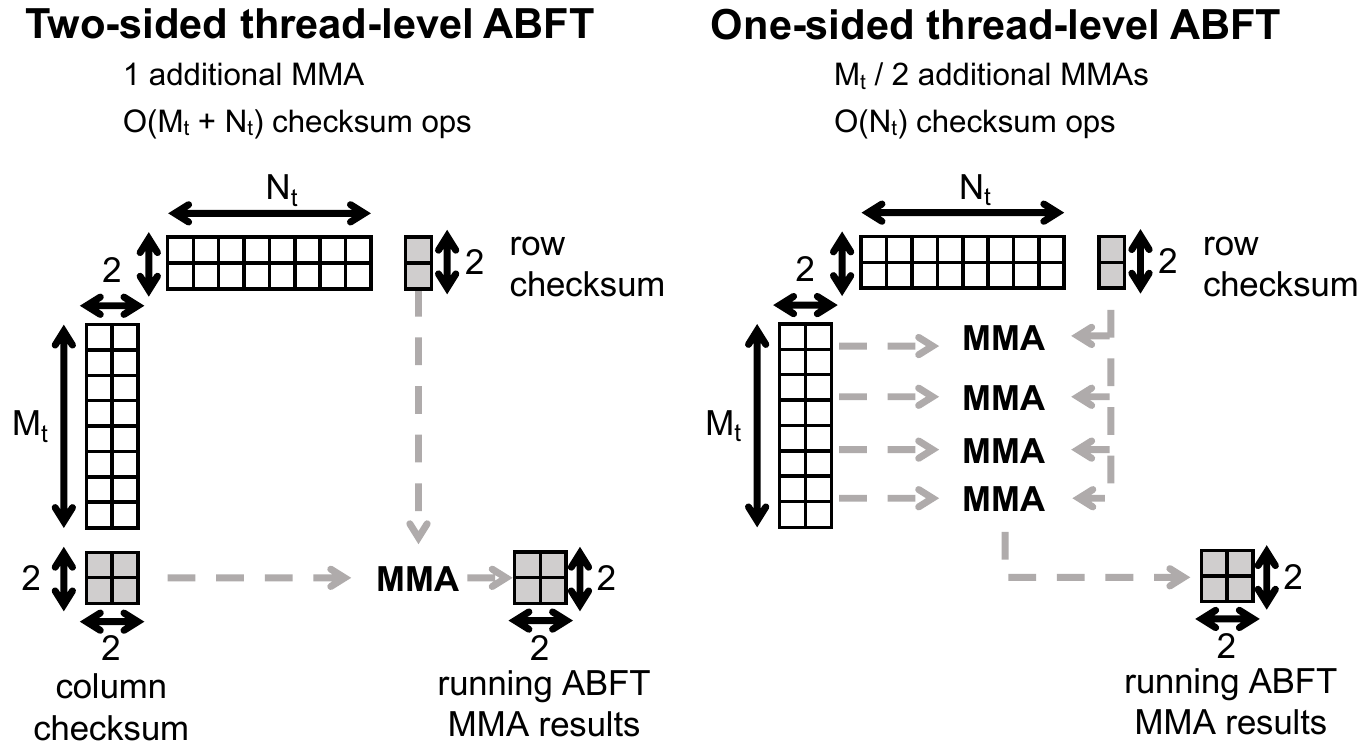}
    \caption{Comparison of two-sided and one-sided  \threadABFT for a single step in the $\mathbf{\gemmK}$ dimension.}
    \label{fig:two_v_one_sided}
\end{figure}

Adopting the \abft approach described in \Section\ref{sec:background:abft} at thread level would involve performing the following for each step the thread takes along the $\gemmK$ dimension: (1)~computing a thread-level \actChecksum from $\matAThread$, (2)~computing a thread-level \weightChecksum from $\matBThread$, and (3)~performing a single \mma over these checksums to generate \abft output values. These steps are illustrated in the left-hand side of \Figure\ref{fig:two_v_one_sided}, and are repeated for each iteration along the $\gemmK$ dimension, accumulating into the same \abft output registers. Once the thread has completed all iterations along the $\gemmK$ dimension, it generates a thread-local \checksumC and compares it to the final \abft output registers. We call this approach \textit{two-sided \threadABFT}, as it generates checksums for both $\matAThread$ and $\matBThread$. 

\newcommand{\twoSidedChecksumOps}{$(2\gemmMThread + 2\gemmNThread)$\xspace}
\newcommand{\twoSidedMMAOps}{1\xspace}
Two-sided \threadABFT minimizes the number of redundant \mma operations performed by \threadABFT, as it performs only one extra \mma for every step along the $\gemmK$ dimension. However, it maximizes the amount of computation performed in generating thread-local activation and weight checksums. 

It is important to note that checksum generation involves summations that will execute on traditional arithmetic units on the GPU (\eg using \texttt{HADD2} PTX instructions), rather than on \tensorcores. In contrast, redundant \mma operations will execute on \tensorcores. Thus two-sided \threadABFT will more significantly utilize traditional arithmetic units than \tensorcores because it performs $\mathcal{O}(\gemmMThread + \gemmNThread)$ additional checksum generation operations but only one additional \mma per step along the $\gemmK$ dimension. 

Given that \tensorcores are the drivers behind the math performed in the \matmatmultShort for a \linearlayer, it is \tensorcores that are heavily underutilized by bandwidth-bound \linearlayers, rather than traditional arithmetic units. Traditional arithmetic units are likely not as underutilized in bandwidth-bound \linearlayers, as they are also used by threads to carry out general control flow (\eg updating loop counters) and to assist in loading/storing data (\eg computing addresses). Thus, minimizing the number of additional \mmas performed in two-sided \threadABFT may insufficiently exploit underutilized \tensorcores. At the same time, our experience with replication in \Section\ref{sec:thread:replication} indicates that adding too many additional \mmas can also lead to high overhead. 

\begin{table}[t]
	\centering
	\caption{Additional \tensorcore \mmas and checksum operations done by thread-level replication (Rep.), two-sided \abft, and one-sided \abft per step in the $\mathbf{\gemmK}$ dimension.}
	\begin{tabular}[t]{r|ccc}
    	& \textbf{Rep.} & \textbf{Two-sided} & \textbf{One-sided} \\ \hline
        Tensor Core \mmas & $\gemmMThread\gemmNThread/2$ & 1 & $\gemmMThread/2$ \\
        Checksum ops. & 0 & $\mathcal{O}(\gemmMThread + \gemmNThread)$ & $\mathcal{O}(\gemmNThread)$
	\end{tabular}
    \label{table:thread:variants}
\end{table}

To straddle this tradeoff between added operations to \tensorcores and added operations to traditional arithmetic units, we leverage a \textit{one-sided \threadABFT scheme}. Rather than computing checksums for both $\matAThread$ and  $\matBThread$ and performing a single \mma across these checksums, one-sided \threadABFT instead generates a checksum only for $\matBThread$ and multiplies the entirety of $\matAThread$ with this checksum.\footnote{One can alternatively multiply a checksum of $\matAThread$ with $\matBThread$. We have selected the converse due to ease of implementation in \cutlass.
} As illustrated in the right-hand side of \Figure\ref{fig:two_v_one_sided} this results in performing $\frac{\gemmMThread}{2}$ additional \mmas, and $\mathcal{O}(\gemmNThread)$ checksum generation operations for each step along the $\gemmK$ dimension. 

As shown in Table~\ref{table:thread:variants}, one-sided \threadABFT sits in the ``sweet spot'' between thread-level replication and two-sided \threadABFT in terms of additional \mmas and checksum operations performed. We illustrate in \Section\ref{sec:eval:square} that this enables one-sided \threadABFT to provide the lowest execution-time overhead among these approaches to thread-level redundant execution. 

\begin{figure*}[t]
    \centering
    \newcommand{\figureheight}{1.12in}
    \newcommand{\figurewidth}{\linewidth}
    \input{figures/eval_all}
    \vspace{-0.1in}
    \caption{Execution-time overhead on all \nns considered. %
    To avoid clutter, error bars are not plotted in this figure, but are plotted in Figures~\ref{fig:eval:conv_normal}, \ref{fig:eval:dlrm}, and~\ref{fig:eval:noscope} for each \nn.}
    \label{fig:eval:all}
    \vspace{-0.1in}
\end{figure*}
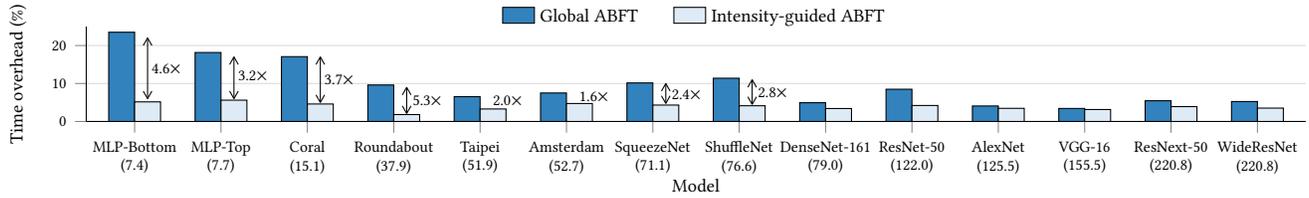

\subsection{Per-layer, intensity-guided adaptation} \label{sec:hybrid} 
As shown in \Section\ref{sec:opp}, \nns have a mix of compute- and bandwidth-bound \linearlayers. The \abft scheme with the lowest execution-time overhead for a given layer depends on the bottleneck of the \layer, with \globalABFT preferable for compute-bound \layers and \threadABFT preferable for bandwidth-bound \layers. 

Rather than selecting one \abft scheme to be applied to all \linearlayers of a \nn, we propose \textit{\hybridABFT}, which selects among \globalABFT and \threadABFT for each individual \linearlayer. Prior to deploying a \nn, \hybridABFT measures the execution-time overhead of each \linearlayer under \globalABFT and \threadABFT, and chooses the scheme with the lowest overhead for that \layer. %
As we show in \Section\ref{sec:eval}, in conforming to the ideas presented in this paper, \linearlayers with higher \aiShort typically benefit from \globalABFT, while those with lower \aiShort typically benefit from \threadABFT. Thus, \hybridABFT uses \aiShort as a guide in selecting the best \abft scheme for each \layer. Our evaluation in \Section\ref{sec:eval} shows that \hybridABFT significantly reduces overhead compared to either global or \threadABFT alone.

\revised{
\textbf{Integration with pre-deployment optimizers.}
\HybridABFT fits alongside the popular approach of pre-deployment optimization in \nn inference, as performed by frameworks like TensorRT~\cite{nvidia-tensorrt}, TVM~\cite{chen2018tvm},  cuDNN~\cite{nvidia-cudnn}, and \cutlass. This process takes in a \nn and an input size (e.g., image resolution, batch size) that will be used during inference  and enumerates and executes all configurations of each  layer  in the \nn (e.g., tile sizes, matrix layouts). The configuration with the lowest execution time for a layer is chosen for that layer and used for all inference requests during deployment. 
A pre-deployment optimizer using \hybridABFT will include \globalABFT and \threadABFT in its enumeration of configurations of a \matmatmult. \HybridABFT chooses the fastest among these, which typically aligns with the \ai of the layer, as we show in \Section\ref{sec:eval}.%
}

%% file: figures/eval_all.tex
\begin{tikzpicture}[]
\pgfplotsset{label style={font=\footnotesize}, 
             tick label style={font=\scriptsize},
             legend style={font=\footnotesize},
             every non boxed x axis/.append style={x axis line style=-},
             every non boxed y axis/.append style={y axis line style=-},
             axis lines=left,
             /pgfplots/ybar legend/.style={
                /pgfplots/legend image code/.code={%
                    \draw[##1,/tikz/.cd,yshift=-0.25em]
                    (0cm,0cm) rectangle (12pt,6pt);
                },
            },
        }

\begin{axis}[
height=\figureheight,
legend cell align={left},
legend columns=2,
legend style={at={(0.5,1.3)}, anchor=north, draw=none, /tikz/every even column/.append style={column sep=0.4cm}},
tick align=outside,
tick label style={/pgf/number format/assume math mode},
tick pos=left,
width=\figurewidth,
x grid style={white!69.01960784313725!black},
xmin=-0.41, xmax=13.71,
xticklabel style={align=center},
xtick={0.150,1.150,2.150,3.150,4.150,5.150,6.150,7.150,8.150,9.150,10.150,11.150,12.150,13.150},
xticklabels={
{\nameDLRMBottom\\(\aiBatchOneDLRMBottom)},
{\nameDLRMTop\\(\aiBatchOneDLRMTop)},
{\shortNameNoscopeCoral\\(\aiBatchSixtyFourNoscopeCoral)},
{\shortNameNoscopeRoundabout\\(\aiBatchSixtyFourNoscopeRoundabout)},
{\shortNameNoscopeTaipei\\(\aiBatchSixtyFourNoscopeTaipei)},
{\shortNameNoscopeAmsterdam\\(\aiBatchSixtyFourNoscopeAmsterdam)},
{\nameSqueezenet\\(\aiBatchOneResBigSqueezenet)},
{\nameShufflenet\\(\aiBatchOneResBigShufflenet)},
{\nameDensenet\\(\aiBatchOneResBigDensenet)},
{\nameResnet\\(\aiBatchOneResBigResnet)},
{\nameAlexnet\\(\aiBatchOneResBigAlexnet)},
{\nameVgg\\(\aiBatchOneResBigVgg)},
{\nameResnext\\(\aiBatchOneResBigResnext)},
{\nameWideResnetShort\\(\aiBatchOneResBigWideResnet)}},
xlabel style={align=center,at={(axis description cs:0.5,-.1)}},
xlabel={Model},
y grid style={gray, opacity=0.3},
ylabel style={at={(axis description cs:0.02,.5)}},
ymajorgrids,
ylabel style={align=left},
ylabel={\SlowdownMetric (\%)},
ymin=0, ymax=25
]

\addlegendimage{ybar,ybar legend,draw=black,fill=\colorGlobal};
\addlegendentry{\GlobalABFT}

\addlegendimage{ybar,ybar legend,draw=black,fill=\colorHybrid};
\addlegendentry{\HybridABFT}

\draw[draw=black,fill=\colorGlobal] (axis cs:-0.150,0) rectangle (axis cs:0.150,\globalOverheadBatchOneDLRMBottom);
\draw[draw=black,fill=\colorHybrid] (axis cs:0.150,0) rectangle (axis cs:0.450,\hybridOverheadBatchOneDLRMBottom);

\draw[draw=black,fill=\colorGlobal] (axis cs:0.850,0) rectangle (axis cs:1.150,\globalOverheadBatchOneDLRMTop);
\draw[draw=black,fill=\colorHybrid] (axis cs:1.150,0) rectangle (axis cs:1.450,\hybridOverheadBatchOneDLRMTop);

\draw[draw=black,fill=\colorGlobal] (axis cs:1.850,0) rectangle (axis cs:2.150,\globalOverheadBatchSixtyFourNoscopeCoral);
\draw[draw=black,fill=\colorHybrid] (axis cs:2.150,0) rectangle (axis cs:2.450,\hybridOverheadBatchSixtyFourNoscopeCoral);

\draw[draw=black,fill=\colorGlobal] (axis cs:2.850,0) rectangle (axis cs:3.150,\globalOverheadBatchSixtyFourNoscopeRoundabout);
\draw[draw=black,fill=\colorHybrid] (axis cs:3.150,0) rectangle (axis cs:3.450,\hybridOverheadBatchSixtyFourNoscopeRoundabout);

\draw[draw=black,fill=\colorGlobal] (axis cs:3.850,0) rectangle (axis cs:4.150,\globalOverheadBatchSixtyFourNoscopeTaipei);
\draw[draw=black,fill=\colorHybrid] (axis cs:4.150,0) rectangle (axis cs:4.450,\hybridOverheadBatchSixtyFourNoscopeTaipei);

\draw[draw=black,fill=\colorGlobal] (axis cs:4.850,0) rectangle (axis cs:5.150,\globalOverheadBatchSixtyFourNoscopeAmsterdam);
\draw[draw=black,fill=\colorHybrid] (axis cs:5.150,0) rectangle (axis cs:5.450,\hybridOverheadBatchSixtyFourNoscopeAmsterdam);

\draw[draw=black,fill=\colorGlobal] (axis cs:5.850,0) rectangle (axis cs:6.150,\globalOverheadBatchOneResBigSqueezenet);
\draw[draw=black,fill=\colorHybrid] (axis cs:6.150,0) rectangle (axis cs:6.450,\hybridOverheadBatchOneResBigSqueezenet);

\draw[draw=black,fill=\colorGlobal] (axis cs:6.850,0) rectangle (axis cs:7.150,\globalOverheadBatchOneResBigShufflenet);
\draw[draw=black,fill=\colorHybrid] (axis cs:7.150,0) rectangle (axis cs:7.450,\hybridOverheadBatchOneResBigShufflenet);

\draw[draw=black,fill=\colorGlobal] (axis cs:7.850,0) rectangle (axis cs:8.150,\globalOverheadBatchOneResBigDensenet);
\draw[draw=black,fill=\colorHybrid] (axis cs:8.150,0) rectangle (axis cs:8.450,\hybridOverheadBatchOneResBigDensenet);

\draw[draw=black,fill=\colorGlobal] (axis cs:8.850,0) rectangle (axis cs:9.150,\globalOverheadBatchOneResBigResnet);
\draw[draw=black,fill=\colorHybrid] (axis cs:9.150,0) rectangle (axis cs:9.450,\hybridOverheadBatchOneResBigResnet);

\draw[draw=black,fill=\colorGlobal] (axis cs:9.850,0) rectangle (axis cs:10.150,\globalOverheadBatchOneResBigAlexnet);
\draw[draw=black,fill=\colorHybrid] (axis cs:10.150,0) rectangle (axis cs:10.450,\hybridOverheadBatchOneResBigAlexnet);

\draw[draw=black,fill=\colorGlobal] (axis cs:10.850,0) rectangle (axis cs:11.150,\globalOverheadBatchOneResBigVgg);
\draw[draw=black,fill=\colorHybrid] (axis cs:11.150,0) rectangle (axis cs:11.450,\hybridOverheadBatchOneResBigVgg);

\draw[draw=black,fill=\colorGlobal] (axis cs:11.850,0) rectangle (axis cs:12.150,\globalOverheadBatchOneResBigResnext);
\draw[draw=black,fill=\colorHybrid] (axis cs:12.150,0) rectangle (axis cs:12.450,\hybridOverheadBatchOneResBigResnext);

\draw[draw=black,fill=\colorGlobal] (axis cs:12.850,0) rectangle (axis cs:13.150,\globalOverheadBatchOneResBigWideResnet);
\draw[draw=black,fill=\colorHybrid] (axis cs:13.150,0) rectangle (axis cs:13.450,\hybridOverheadBatchOneResBigWideResnet);

\node[anchor=west] at (axis cs: 0.25,14) {\scriptsize 4.6$\times$};
\addplot [thin, black, solid]
table {%
0.3 6
0.3 22
};
\addplot [thin, black, solid]
table {%
0.3 6
0.35 8
};
\addplot [thin, black, solid]
table {%
0.3 6
0.25 8
};
\addplot [thin, black, solid]
table {%
0.3 22
0.35 20
};
\addplot [thin, black, solid]
table {%
0.3 22
0.25 20
};

\node[anchor=west] at (axis cs: 1.25,12) {\scriptsize 3.2$\times$};
\addplot [thin, black, solid]
table {%
1.3 6
1.3 17
};
\addplot [thin, black, solid]
table {%
1.3 6
1.35 8
};
\addplot [thin, black, solid]
table {%
1.3 6
1.25 8
};
\addplot [thin, black, solid]
table {%
1.3 17
1.35 15
};
\addplot [thin, black, solid]
table {%
1.3 17
1.25 15
};

\node[anchor=west] at (axis cs: 2.25,11) {\scriptsize 3.7$\times$};
\addplot [thin, black, solid]
table {%
2.3 5
2.3 17
};
\addplot [thin, black, solid]
table {%
2.3 5
2.35 7
};
\addplot [thin, black, solid]
table {%
2.3 5
2.25 7
};
\addplot [thin, black, solid]
table {%
2.3 17
2.35 15
};
\addplot [thin, black, solid]
table {%
2.3 17
2.25 15
};

\node[anchor=west] at (axis cs: 3.27,5.5) {\scriptsize 5.3$\times$};
\addplot [thin, black, solid]
table {%
3.3 2.1
3.3 9
};
\addplot [thin, black, solid]
table {%
3.3 2.1
3.35 4.1
};
\addplot [thin, black, solid]
table {%
3.3 2.1
3.25 4.1
};
\addplot [thin, black, solid]
table {%
3.3 9
3.35 7
};
\addplot [thin, black, solid]
table {%
3.3 9
3.25 7
};

\node[anchor=west] at (axis cs: 4.2,5.5) {\scriptsize 2.0$\times$};
\node[anchor=west] at (axis cs: 5.2,6.75) {\scriptsize 1.6$\times$};

\node[anchor=west] at (axis cs: 6.27,7.25) {\scriptsize 2.4$\times$};
\addplot [thin, black, solid]
table {%
6.3 4.75
6.3 10
};
\addplot [thin, black, solid]
table {%
6.3 4.75
6.35 6.75
};
\addplot [thin, black, solid]
table {%
6.3 4.75
6.25 6.75
};
\addplot [thin, black, solid]
table {%
6.3 10
6.35 8
};
\addplot [thin, black, solid]
table {%
6.3 10
6.25 8
};

\node[anchor=west] at (axis cs: 7.27,7.75) {\scriptsize 2.8$\times$};
\addplot [thin, black, solid]
table {%
7.3 4.5
7.3 11
};
\addplot [thin, black, solid]
table {%
7.3 4.5
7.35 6.5
};
\addplot [thin, black, solid]
table {%
7.3 4.5
7.25 6.5
};
\addplot [thin, black, solid]
table {%
7.3 11
7.35 9
};
\addplot [thin, black, solid]
table {%
7.3 11
7.25 9
};

\end{axis}

\end{tikzpicture}

%% file: evaluation.tex
\section{Implementation and evaluation} \label{sec:eval}
We now evaluate the execution-time overhead of \hybridABFT. The highlights of the evaluation are as follows:
\begin{denseitemize}
\item Across eight popular \cnns, two \nns  used in \dlrms, and four specialized \cnns, \hybridABFT reduces execution-time overhead compared to \globalABFT by 1.09--5.3$\times$. %
\item \HybridABFT provides the largest reductions in execution-time overhead for \nns that have many \linearlayers with low \aiShort, such as \dlrms (up to 4.9$\times$ reduction) and specialized \cnns (up to 5.3$\times$ reduction).
\item Even for \nns that have many \linearlayers with high \aiShort, \hybridABFT still significantly reduces execution-time overhead (\eg 1.5$\times$ for \nameWideResnet). This shows the benefit of \hybridABFT's adaptive approach to \abft, as even \nns that are primarily compute bound often have some \linearlayers with low \ai.
\item \HybridABFT provides similar benefits across various input resolutions (\Section\ref{sec:eval:general_cnns}) and batch sizes (\Section\ref{sec:eval:dlrm}).
\item The one-sided \threadABFT approach motivated in \Section\ref{sec:thread:checksum} significantly reduces execution-time overhead compared to two-sided \threadABFT and thread-level replication (\Section\ref{sec:eval:square}).
\end{denseitemize}

\subsection{Implementation} \label{sec:implementation}
Recall that \hybridABFT adapts to each \linearlayer in a \nn by choosing between \threadABFT and \globalABFT. We implement \threadABFT and \globalABFT in CUDA/C++ atop \cutlass~\cite{nvidia-cutlass}, a high-performance, open-source \matmatmult library developed by NVIDIA. For \threadABFT, we modify existing thread-level inner loops in \cutlass to perform checksum generation, redundant \mmas, and final checksum comparison. We implement the \globalABFT scheme based on the state-of-the-art approach from \harietal (discussed in \Section\ref{sec:background:abft_nn}), using NVIDIA's CUB library~\cite{nvidia-cub} when possible. 

\revised{{Recall from \Section\ref{sec:hybrid} that \hybridABFT fits alongside common pre-deployment \nnShort optimizers.} We integrate \hybridABFT into the pre-deployment workflow of the \cutlass profiler, which selects the fastest \matmatmult kernel and configuration (e.g., tile size, layout) for a given \matmatmult size.} %

\subsection{Evaluation setup} \label{sec:eval:setup}
\textbf{Baselines.} Our main comparison is between \hybridABFT and the state-of-the-art approach to \globalABFT for \nn inference on GPUs described in \Section\ref{sec:background:abft_nn}. We also evaluate one-sided \threadABFT alone (referred to as ``\threadABFT''), and in \Section\ref{sec:eval:square} compare to two-sided \threadABFT and thread-level replication.

\begin{figure*}[t]
    \centering
    \newcommand{\figureheight}{1.05in}
    \newcommand{\figurewidth}{\linewidth}
    \input{figures/eval_conv_normal}
    \vspace{-0.1in}
    \caption{Execution-time overhead on general-purpose \cnns with inputs of resolution $\mathbf{1080\times1920}$. %
    }
    \label{fig:eval:conv_normal}
\end{figure*}
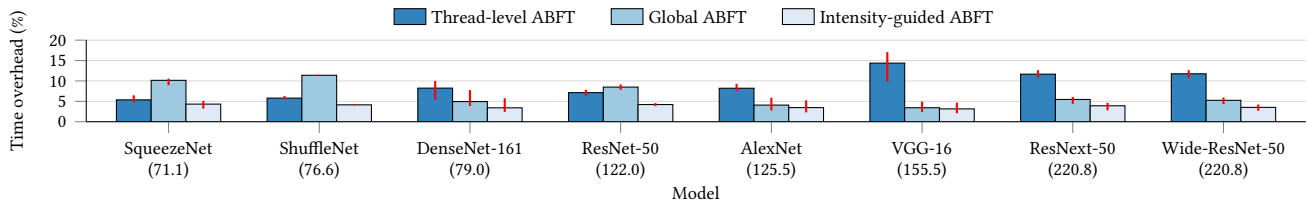

\begin{figure}[t]
    \centering
    \newcommand{\figureheight}{1.05in}
    \newcommand{\figurewidth}{\linewidth}
    \input{figures/eval_dlrm}
    \vspace{-0.3in}
    \caption{Execution-time overheads on \nns from \dlrm. %
    Error bars are tight to the point of being imperceptible.
    }
    \label{fig:eval:dlrm}
\end{figure}
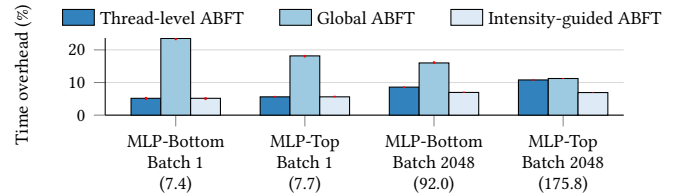

\newcommand{\exeTime}{T}
\newcommand{\exeTimeOg}{\exeTime_{o}\xspace}
\newcommand{\exeTimeFt}{\exeTime_{r}\xspace}
\textbf{Metrics.} Execution-time overhead is one of the primary metrics of interest for redundant execution. For each \linearlayer of a \nn, we obtain the execution time of the original \matmatmultShort without redundancy ($\exeTimeOg$), as well as that of the redundant version ($\exeTimeFt$) and report the percentage increase in execution time ($\frac{\exeTimeFt - \exeTimeOg}{\exeTimeOg} * 100$). We report execution-time overhead for an entire \nn by summing the per-\layer execution times and using these in the equation above. {We include only \linearlayers, as these layers typically dominate the end-to-end execution time of a \nn.} {Moreover, aggregating the  execution times of each \linearlayer in this fashion is representative of overall execution-time overhead for the \nn as a whole, because, for all of the \nns we consider, the subsequent layer of the \nn cannot begin executing until the current layer has completed execution.} We report the mean of 10 trials of 1000 runs after 100 warmup runs. Error bars show the maximum and minimum time overheads across trials. In many cases, error bars are imperceptible due to their tightness. We do not plot error bars in \Figure\ref{fig:eval:all} to avoid clutter and because all error bars for \nns in \Figure\ref{fig:eval:all} are plotted in \Figures\ref{fig:eval:conv_normal}, \ref{fig:eval:dlrm}, and~\ref{fig:eval:noscope}. Note that while in some cases error bars may give the incorrect impression that \hybridABFT performs worse than \globalABFT, \hybridABFT, by design, always performs at least as well as \globalABFT.

We also report ``\aggregateAI'' (defined in \Section\ref{sec:opp:ai}). \updated{In each figure, the FP16 \aggregateAI is listed in parentheses below each model.}

\textbf{Evaluation setting.} We evaluate on an NVIDIA T4 GPU~\cite{nvidia-turing}, which is a state-of-the-art inference-optimized GPU, on an AWS g4dn.xlarge instance. The T4 offers \tFourHalfTFLOPS FP16 \TFLOPS and \tFourGBps GB/sec of memory bandwidth, giving it an FP16 \cmr of \tFourHalfCMR. We use CUDA 11.0 and configure the clockrate of the GPU according to that used in \cutlass~\cite{nvidia-cutlass-clock}. We perform all experiments using FP16 datatypes and we use the \mSixteenNeightKeight \matmatmultsShort targeting \tensorcores described in \Section\ref{sec:background:mm}. Note that it is standard to perform \nn inference in low precision, such as FP16. We pad matrix dimensions $\gemmM$, $\gemmN$, and $\gemmK$ to be multiples of eight when needed to operate with the \mSixteenNeightKeight operation.  We find \cutlass's \mSixteenNeightKeight \matmatmultShort with $\gemmM = \gemmN = \gemmK = 2048$ to achieve similar \TFLOPS to the highest reported on the T4 GPU~\cite{jia2019dissecting}. 

\textbf{Workloads.} We consider workloads from multiple domains: %

\textit{General-purpose \cnns.} We consider eight widely-used \cnns from the popular PyTorch Torchvision library~\cite{pytorch-torchvision-cnns}: \nameResnet, \nameVgg, \nameAlexnet, \nameSqueezenet, \nameShufflenet, \nameDensenet, \nameResnext, and \nameWideResnet. Each of these \cnns has 1000 output classes, as is standard for ImageNet. We primarily report performance when operating over HD images of size $1080\times1920$ with batch size of one, though we consider other image resolutions in \Section\ref{sec:eval:general_cnns}.

\textit{Recommendation models.} We consider Facebook's \dlrm~\cite{naumov2019deep}, which has two \nns consisting of fully-connected \layers (also called multilayer perceptrons, or MLPs): \nameDLRMBottom, which has three hidden \layers with 512, 256, and 64 nodes each, and \nameDLRMTop which has two hidden \layers with 512 and 256 nodes, and produces one output value. We primarily consider \dlrms with batch size of one as this is the common case for low-latency, user-facing inference~\cite{zhang2018deepcpu,chung2018serving}. For completeness, we also consider large batch size in \Section\ref{sec:eval:dlrm}. 

\textit{Specialized \cnns.} We also evaluate on \nns representative of ongoing efforts to deploy small \nns (described in \Section\ref{sec:opp:small_nn}). We consider four specialized \cnns used within the NoScope system~\cite{kang2017noscope}: \shortNameNoscopeCoral, \shortNameNoscopeRoundabout, \shortNameNoscopeTaipei, \shortNameNoscopeAmsterdam. These \cnns act as lightweight filters performing binary classification in front of large, general-purpose \cnns for high-throughput offline video analytics in cluster settings. These \cnns have 2--4 convolutional \layers, each with 16--64 channels, at most two fully-connected \layers, and operate over regions of video frames of size $50\times50$ pixels. As these \cnns are used for offline analytics, we use a large batch size of 64 for experiments. %

\textit{Square \matmatmults.} We finally perform a more detailed comparison of one-sided \threadABFT and \globalABFT, along with two-sided \threadABFT and thread-level replication on \matmatmultsShort with $\gemmM = \gemmN = \gemmK$ of various sizes (\Section\ref{sec:eval:square}). 

\subsection{Summary of results} \label{sec:eval:summary}
\Figure\ref{fig:eval:all} compares the execution-time overhead of \globalABFT to that of \hybridABFT on all \nns we consider (listed in order of increasing \aggregateAI).\footnote{{Note that the execution-time overheads do not monotonically decrease with increasing \aggregateAI since execution is performed \layer-wise whereas the \aggregateAI metric is not \layer-wise.}}  Compared to \globalABFT, \hybridABFT reduces execution-time overhead by up to 5.3$\times$. For example, for the \shortNameNoscopeCoral specialized \cnn, \hybridABFT reduces execution-time overhead from 17\% to 4.6\%. As expected, \hybridABFT achieves the largest reduction in execution-time overhead for \nns with low \aggregateAI, as these \nns contain more bandwidth-bound \linearlayers that benefit from \threadABFT. That said, \hybridABFT reduces execution-time overhead considerably even for \nns with high \aggregateAI. For example, \hybridABFT reduces the execution-time overhead on \nameWideResnet by 1.5$\times$ compared to \globalABFT (from 5.3\% to 3.5\%). Even though such \nns have high aggregate \aiShort, they still contain bandwidth-bound \linearlayers, for which using \threadABFT 
over \globalABFT reduces overhead. 

\subsection{Evaluation across various \nn domains} 
\subsubsection{General-purpose \cnns} \label{sec:eval:general_cnns}
\Figure\ref{fig:eval:conv_normal} shows the execution-time overhead for \threadABFT, \globalABFT, and \hybridABFT on eight popular general-purpose \cnns operating over HD images of size $1080\times1920$ at batch size one. Compared to \globalABFT, \hybridABFT reduces execution-time overhead by 1.09--2.75$\times$. As expected, \threadABFT obtains lower execution-time overhead than \globalABFT for \cnns with low \aggregateAI, while \globalABFT has lower overhead for \cnns with higher \aggregateAI. \HybridABFT obtains the lowest execution-time overhead across all the \cnns, motivating its per-\layer, \aiShort-guided approach. 

\textit{Effect of image resolution.} 
When operating on images of size $224\times224$ (the standard resolution in ImageNet), \hybridABFT reduces execution-time overhead by 1.3--3.3$\times$ compared to \globalABFT. 
This larger reduction compared to operating on HD images stems from the lower \aggregateAI of \cnns when operating on images with smaller resolution (described in \Section\ref{sec:opp:ai}). This leads to more \linearlayers being bandwidth-bound and benefiting from \threadABFT in \hybridABFT. %

\subsubsection{Recommendation models (\DLRM)} \label{sec:eval:dlrm}
We next consider the \nns used in Facebook's \dlrm. \Figure\ref{fig:eval:dlrm} plots execution-time overheads on \nameDLRMBottom and \nameDLRMTop. %
At batch size of one, which corresponds to low-latency deployments of \dlrms for user-facing services, both \nameDLRMBottom and \nameDLRMTop have low \aggregateAI. This results in \hybridABFT reducing execution-time overhead compared to \globalABFT by 4.55$\times$ and  3.24$\times$ for \nameDLRMBottom and \nameDLRMTop, respectively. At a very large batch size of 2048, the \aggregateAI of both \nameDLRMBottom and \nameDLRMTop increase, but at different rates. The \aggregateAI of \nameDLRMTop increases from 7.7 to 175.8, resulting in the difference between global and thread-level \abft decreasing. In contrast, the \aggregateAI of \nameDLRMBottom grows only from 7.4 to 92, resulting in \threadABFT continuing to have lower overhead. In both cases, \hybridABFT achieves the lowest overhead, illustrating the need for \abft to consider the resource bottlenecks of each \linearlayer of a \nn. 

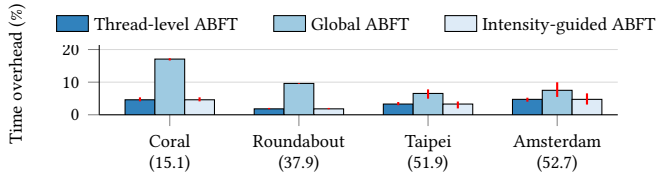
\begin{figure}[t]
    \centering
    \newcommand{\figureheight}{1.05in}
    \newcommand{\figurewidth}{\linewidth}
    \input{figures/eval_conv_noscope}
    \vspace{-0.3in}
    \caption{Execution-time overheads on specialized \cnns. %
    }
    \label{fig:eval:noscope}
\end{figure}

\subsubsection{Specialized \cnns} \label{sec:eval:specialized}
\Figure\ref{fig:eval:noscope} shows the execution-time overheads on specialized \cnns from NoScope~\cite{kang2017noscope} at batch size 64. For these primarily bandwidth-bound \cnns with low \aggregateAI, \hybridABFT reduces execution-time overhead by 1.6--5.3$\times$. These results are particularly promising when considering the growing trends described in \Section\ref{sec:opp} of designing lightweight \nns, coupled with the increasing \cmr of GPUs, which will likely result in more \nns being bandwidth-bound. %

\subsection{Evaluation of thread-level design decisions} \label{sec:eval:square} 
To evaluate the design decisions made in leveraging \threadABFT for \nn inference on GPUs, we now evaluate \globalABFT and the various approaches to thread-level redundant execution described in \Section\ref{sec:thread:replication} and \Section\ref{sec:thread}. We perform such evaluation on square \matmatmults (i.e., $\gemmM = \gemmN = \gemmK$) of varying size, allowing us to control \aiShort and best illustrate the tradeoffs.

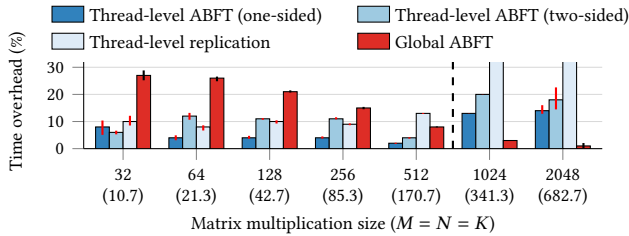
\begin{figure}[t]
    \centering
    \newcommand{\figureheight}{1.12in}
    \newcommand{\figurewidth}{\linewidth} 
    \input{figures/eval_square}
    \vspace{-0.3in}
    \caption{Execution-time overhead on square \matmatmults. %
    Sizes left of the dashed line have \aiShort below the T4's FP16 \cmr. The overhead for replication is above 70\% for the final two sizes, and thus is cut off. 
    }
    \label{fig:eval:square}
\end{figure}

\Figure\ref{fig:eval:square} shows the execution-time overhead of each approach with $\gemmM = \gemmN = \gemmK$ ranging from 32 to 2048, corresponding to FP16 \aisShort of 10 to 683. We first compare only the final version of \threadABFT we leverage (one-sided) to \globalABFT. As expected, for matrix sizes with \aiShort less than the FP16 \cmr of the T4 (\tFourHalfCMR), \threadABFT achieves an execution-time overhead up to 6.5$\times$ lower than that of \globalABFT, while for matrix sizes with higher \aiShort, \globalABFT achieves overheads up to 14$\times$ lower than \threadABFT. It is clear that taking a one-size-fits-all approach to \abft will lead to suboptimal performance on certain matrix sizes, motivating our adaptive approach in \hybridABFT. 

\Figure\ref{fig:eval:square} also shows that one-sided \threadABFT almost always exhibits lower execution-time overhead than two-sided \threadABFT and thread-level replication. This reinforces our decision to use one-sided \abft for \threadABFT. The differences between replication and \abft are particularly stark for larger sizes (512 and beyond), where the overhead of replication sharply spikes due to increasing competition for \tensorcores.

%% file: figures/eval_conv_normal.tex
\begin{tikzpicture}[]
\pgfplotsset{label style={font=\footnotesize}, 
             tick label style={font=\footnotesize},
             legend style={font=\footnotesize},
             every non boxed x axis/.append style={x axis line style=-},
             every non boxed y axis/.append style={y axis line style=-},
             axis lines=left,
             /pgfplots/ybar legend/.style={
                /pgfplots/legend image code/.code={%
                    \draw[##1,/tikz/.cd,yshift=-0.25em]
                    (0cm,0cm) rectangle (12pt,6pt);
                },
            },
        }

\begin{axis}[
height=\figureheight,
legend cell align={left},
legend columns=3,
legend style={at={(0.5,1.5)}, anchor=north, draw=none, /tikz/every even column/.append style={column sep=0.4cm}},
tick align=outside,
tick label style={/pgf/number format/assume math mode},
tick pos=left,
width=\figurewidth,
x grid style={white!69.01960784313725!black},
xmin=-0.41, xmax=10.11,
xticklabel style={align=center},
xtick={0.3,1.6,2.9,4.2,5.5,6.8,8.1,9.4},
xticklabels={
{\nameSqueezenet\\(\aiBatchOneResBigSqueezenet)},
{\nameShufflenet\\(\aiBatchOneResBigShufflenet)},
{\nameDensenet\\(\aiBatchOneResBigDensenet)},
{\nameResnet\\(\aiBatchOneResBigResnet)},
{\nameAlexnet\\(\aiBatchOneResBigAlexnet)},
{\nameVgg\\(\aiBatchOneResBigVgg)},
{\nameResnext\\(\aiBatchOneResBigResnext)},
{\nameWideResnet\\(\aiBatchOneResBigWideResnet)}},
xlabel style={align=center,at={(axis description cs:0.5,-.2)}},
xlabel={Model}, %
y grid style={gray, opacity=0.3},
ylabel style={at={(axis description cs:0.02,.5)}},
ymajorgrids,
ylabel style={align=left},
ylabel={\SlowdownMetric (\%)},
ymin=0, ymax=20
]

\addlegendimage{ybar,ybar legend,draw=black,fill=\colorOgThread};
\addlegendentry{\ThreadABFT}

\addlegendimage{ybar,ybar legend,draw=black,fill=\colorOgGlobal};
\addlegendentry{\GlobalABFT}

\addlegendimage{ybar,ybar legend,draw=black,fill=\colorOgHybrid};
\addlegendentry{\HybridABFT}

\draw[draw=black,fill=\colorOgThread] (axis cs:-0.15,0) rectangle (axis cs:0.15,\threadOverheadBatchOneResBigSqueezenet);
\draw[draw=black,fill=\colorOgGlobal] (axis cs:0.15,0) rectangle (axis cs:0.45,\globalOverheadBatchOneResBigSqueezenet);
\draw[draw=black,fill=\colorOgHybrid] (axis cs:0.45,0) rectangle (axis cs:0.75,\hybridOverheadBatchOneResBigSqueezenet);

\draw[draw=black,fill=\colorOgThread] (axis cs:1.15,0) rectangle (axis cs:1.45,\threadOverheadBatchOneResBigShufflenet);
\draw[draw=black,fill=\colorOgGlobal] (axis cs:1.45,0) rectangle (axis cs:1.75,\globalOverheadBatchOneResBigShufflenet);
\draw[draw=black,fill=\colorOgHybrid] (axis cs:1.75,0) rectangle (axis cs:2.05,\hybridOverheadBatchOneResBigShufflenet);

\draw[draw=black,fill=\colorOgThread] (axis cs:2.45,0) rectangle (axis cs:2.75,\threadOverheadBatchOneResBigDensenet);
\draw[draw=black,fill=\colorOgGlobal] (axis cs:2.75,0) rectangle (axis cs:3.05,\globalOverheadBatchOneResBigDensenet);
\draw[draw=black,fill=\colorOgHybrid] (axis cs:3.05,0) rectangle (axis cs:3.35,\hybridOverheadBatchOneResBigDensenet);

\draw[draw=black,fill=\colorOgThread] (axis cs:3.75,0) rectangle (axis cs:4.05,\threadOverheadBatchOneResBigResnet);
\draw[draw=black,fill=\colorOgGlobal] (axis cs:4.05,0) rectangle (axis cs:4.35,\globalOverheadBatchOneResBigResnet);
\draw[draw=black,fill=\colorOgHybrid] (axis cs:4.35,0) rectangle (axis cs:4.65,\hybridOverheadBatchOneResBigResnet);

\draw[draw=black,fill=\colorOgThread] (axis cs:5.05,0) rectangle (axis cs:5.35,\threadOverheadBatchOneResBigAlexnet);
\draw[draw=black,fill=\colorOgGlobal] (axis cs:5.35,0) rectangle (axis cs:5.65,\globalOverheadBatchOneResBigAlexnet);
\draw[draw=black,fill=\colorOgHybrid] (axis cs:5.65,0) rectangle (axis cs:5.95,\hybridOverheadBatchOneResBigAlexnet);

\draw[draw=black,fill=\colorOgThread] (axis cs:6.35,0) rectangle (axis cs:6.65,\threadOverheadBatchOneResBigVgg);
\draw[draw=black,fill=\colorOgGlobal] (axis cs:6.65,0) rectangle (axis cs:6.95,\globalOverheadBatchOneResBigVgg);
\draw[draw=black,fill=\colorOgHybrid] (axis cs:6.95,0) rectangle (axis cs:7.25,\hybridOverheadBatchOneResBigVgg);

\draw[draw=black,fill=\colorOgThread] (axis cs:7.65,0) rectangle (axis cs:7.95,\threadOverheadBatchOneResBigResnext);
\draw[draw=black,fill=\colorOgGlobal] (axis cs:7.95,0) rectangle (axis cs:8.25,\globalOverheadBatchOneResBigResnext);
\draw[draw=black,fill=\colorOgHybrid] (axis cs:8.25,0) rectangle (axis cs:8.55,\hybridOverheadBatchOneResBigResnext);

\draw[draw=black,fill=\colorOgThread] (axis cs:8.95,0) rectangle (axis cs:9.25,\threadOverheadBatchOneResBigWideResnet);
\draw[draw=black,fill=\colorOgGlobal] (axis cs:9.25,0) rectangle (axis cs:9.55,\globalOverheadBatchOneResBigWideResnet);
\draw[draw=black,fill=\colorOgHybrid] (axis cs:9.55,0) rectangle (axis cs:9.85,\hybridOverheadBatchOneResBigWideResnet);

\path [draw=\colorErrorBars, thick](axis cs:0,\threadOverheadBatchOneResBigSqueezenetPZero)--(axis cs:0,\threadOverheadBatchOneResBigSqueezenetPOneHundred);
\path [draw=\colorErrorBars, thick](axis cs:0.3,\globalOverheadBatchOneResBigSqueezenetPZero)--(axis cs:0.3,\globalOverheadBatchOneResBigSqueezenetPOneHundred);
\path [draw=\colorErrorBars, thick](axis cs:0.6,\hybridOverheadBatchOneResBigSqueezenetPZero)--(axis cs:0.6,\hybridOverheadBatchOneResBigSqueezenetPOneHundred);

\path [draw=\colorErrorBars, thick](axis cs:1.3,\threadOverheadBatchOneResBigShufflenetPZero)--(axis cs:1.3,\threadOverheadBatchOneResBigShufflenetPOneHundred);
\path [draw=\colorErrorBars, thick](axis cs:1.6,\globalOverheadBatchOneResBigShufflenetPZero)--(axis cs:1.6,\globalOverheadBatchOneResBigShufflenetPOneHundred);
\path [draw=\colorErrorBars, thick](axis cs:1.9,\hybridOverheadBatchOneResBigShufflenetPZero)--(axis cs:1.9,\hybridOverheadBatchOneResBigShufflenetPOneHundred);

\path [draw=\colorErrorBars, thick](axis cs:2.6,\threadOverheadBatchOneResBigDensenetPZero)--(axis cs:2.6,\threadOverheadBatchOneResBigDensenetPOneHundred);
\path [draw=\colorErrorBars, thick](axis cs:2.9,\globalOverheadBatchOneResBigDensenetPZero)--(axis cs:2.9,\globalOverheadBatchOneResBigDensenetPOneHundred);
\path [draw=\colorErrorBars, thick](axis cs:3.2,\hybridOverheadBatchOneResBigDensenetPZero)--(axis cs:3.2,\hybridOverheadBatchOneResBigDensenetPOneHundred);

\path [draw=\colorErrorBars, thick](axis cs:3.9,\threadOverheadBatchOneResBigResnetPZero)--(axis cs:3.9,\threadOverheadBatchOneResBigResnetPOneHundred);
\path [draw=\colorErrorBars, thick](axis cs:4.2,\globalOverheadBatchOneResBigResnetPZero)--(axis cs:4.2,\globalOverheadBatchOneResBigResnetPOneHundred);
\path [draw=\colorErrorBars, thick](axis cs:4.5,\hybridOverheadBatchOneResBigResnetPZero)--(axis cs:4.5,\hybridOverheadBatchOneResBigResnetPOneHundred);

\path [draw=\colorErrorBars, thick](axis cs:5.2,\threadOverheadBatchOneResBigAlexnetPZero)--(axis cs:5.2,\threadOverheadBatchOneResBigAlexnetPOneHundred);
\path [draw=\colorErrorBars, thick](axis cs:5.5,\globalOverheadBatchOneResBigAlexnetPZero)--(axis cs:5.5,\globalOverheadBatchOneResBigAlexnetPOneHundred);
\path [draw=\colorErrorBars, thick](axis cs:5.8,\hybridOverheadBatchOneResBigAlexnetPZero)--(axis cs:5.8,\hybridOverheadBatchOneResBigAlexnetPOneHundred);

\path [draw=\colorErrorBars, thick](axis cs:6.5,\threadOverheadBatchOneResBigVggPZero)--(axis cs:6.5,\threadOverheadBatchOneResBigVggPOneHundred);
\path [draw=\colorErrorBars, thick](axis cs:6.8,\globalOverheadBatchOneResBigVggPZero)--(axis cs:6.8,\globalOverheadBatchOneResBigVggPOneHundred);
\path [draw=\colorErrorBars, thick](axis cs:7.1,\hybridOverheadBatchOneResBigVggPZero)--(axis cs:7.1,\hybridOverheadBatchOneResBigVggPOneHundred);

\path [draw=\colorErrorBars, thick](axis cs:7.8,\threadOverheadBatchOneResBigResnextPZero)--(axis cs:7.8,\threadOverheadBatchOneResBigResnextPOneHundred);
\path [draw=\colorErrorBars, thick](axis cs:8.1,\globalOverheadBatchOneResBigResnextPZero)--(axis cs:8.1,\globalOverheadBatchOneResBigResnextPOneHundred);
\path [draw=\colorErrorBars, thick](axis cs:8.4,\hybridOverheadBatchOneResBigResnextPZero)--(axis cs:8.4,\hybridOverheadBatchOneResBigResnextPOneHundred);

\path [draw=\colorErrorBars, thick](axis cs:9.1,\threadOverheadBatchOneResBigWideResnetPZero)--(axis cs:9.1,\threadOverheadBatchOneResBigWideResnetPOneHundred);
\path [draw=\colorErrorBars, thick](axis cs:9.4,\globalOverheadBatchOneResBigWideResnetPZero)--(axis cs:9.4,\globalOverheadBatchOneResBigWideResnetPOneHundred);
\path [draw=\colorErrorBars, thick](axis cs:9.7,\hybridOverheadBatchOneResBigWideResnetPZero)--(axis cs:9.7,\hybridOverheadBatchOneResBigWideResnetPOneHundred);

\end{axis}

\end{tikzpicture}

%% file: figures/eval_dlrm.tex
\begin{tikzpicture}[]
\pgfplotsset{label style={font=\footnotesize}, 
             tick label style={font=\footnotesize},
             legend style={font=\footnotesize},
             every non boxed x axis/.append style={x axis line style=-},
             every non boxed y axis/.append style={y axis line style=-},
             axis lines=left,
             /pgfplots/ybar legend/.style={
                /pgfplots/legend image code/.code={%
                    \draw[##1,/tikz/.cd,yshift=-0.25em]
                    (0cm,0cm) rectangle (12pt,6pt);
                },
            },
        }

\begin{axis}[
height=\figureheight,
legend cell align={left},
legend columns=3,
legend style={at={(0.5,1.41)}, anchor=north, draw=none, /tikz/every even column/.append style={column sep=0.4cm}},
tick align=outside,
tick label style={/pgf/number format/assume math mode},
tick pos=left,
width=\figurewidth,
x grid style={white!69.01960784313725!black},
xmin=-0.41, xmax=4.81,
xticklabel style={align=center},
xtick={0.3,1.6,2.9,4.2},%
xticklabels={{\nameDLRMBottom\\Batch 1\\(\aiBatchOneDLRMBottom)},{\nameDLRMTop\\Batch 1\\(\aiBatchOneDLRMTop)},{\nameDLRMBottom\\Batch 2048\\(\aiBatchTwentyFourtyEightDLRMBottom)},{\nameDLRMTop\\Batch 2048\\(\aiBatchTwentyFourtyEightDLRMTop)}},
y grid style={gray, opacity=0.3},
ylabel style={at={(axis description cs:0.02,.5)}},
ymajorgrids,
ylabel style={align=left},
ylabel={\SlowdownMetric (\%)},
ymin=0, ymax=25
]

\addlegendimage{ybar,ybar legend,draw=black,fill=color_blue_3_2};
\addlegendentry{\ThreadABFT}

\addlegendimage{ybar,ybar legend,draw=black,fill=color_blue_3_1};
\addlegendentry{\GlobalABFT}

\addlegendimage{ybar,ybar legend,draw=black,fill=color_blue_3_0};
\addlegendentry{\HybridABFT}

\draw[draw=black,fill=color_blue_3_2] (axis cs:-0.15,0) rectangle (axis cs:0.15,\threadOverheadBatchOneDLRMBottom);
\draw[draw=black,fill=color_blue_3_1] (axis cs:0.15,0) rectangle (axis cs:0.45,\globalOverheadBatchOneDLRMBottom);
\draw[draw=black,fill=color_blue_3_0] (axis cs:0.45,0) rectangle (axis cs:0.75,\hybridOverheadBatchOneDLRMBottom);

\draw[draw=black,fill=color_blue_3_2] (axis cs:1.15,0) rectangle (axis cs:1.45,\threadOverheadBatchOneDLRMTop);
\draw[draw=black,fill=color_blue_3_1] (axis cs:1.45,0) rectangle (axis cs:1.75,\globalOverheadBatchOneDLRMTop);
\draw[draw=black,fill=color_blue_3_0] (axis cs:1.75,0) rectangle (axis cs:2.05,\hybridOverheadBatchOneDLRMTop);

\draw[draw=black,fill=color_blue_3_2] (axis cs:2.45,0) rectangle (axis cs:2.75,\threadOverheadBatchTwentyFourtyEightDLRMBottom);
\draw[draw=black,fill=color_blue_3_1] (axis cs:2.75,0) rectangle (axis cs:3.05,\globalOverheadBatchTwentyFourtyEightDLRMBottom);
\draw[draw=black,fill=color_blue_3_0] (axis cs:3.05,0) rectangle (axis cs:3.35,\hybridOverheadBatchTwentyFourtyEightDLRMBottom);

\draw[draw=black,fill=color_blue_3_2] (axis cs:3.75,0) rectangle (axis cs:4.05,\threadOverheadBatchTwentyFourtyEightDLRMTop);
\draw[draw=black,fill=color_blue_3_1] (axis cs:4.05,0) rectangle (axis cs:4.35,\globalOverheadBatchTwentyFourtyEightDLRMTop);
\draw[draw=black,fill=color_blue_3_0] (axis cs:4.35,0) rectangle (axis cs:4.65,\hybridOverheadBatchTwentyFourtyEightDLRMTop);

\path [draw=\colorErrorBars, thick](axis cs:0,\threadOverheadBatchOneDLRMBottomPZero)--(axis cs:0,\threadOverheadBatchOneDLRMBottomPOneHundred);
\path [draw=\colorErrorBars, thick](axis cs:0.3,\globalOverheadBatchOneDLRMBottomPZero)--(axis cs:0.3,\globalOverheadBatchOneDLRMBottomPOneHundred);
\path [draw=\colorErrorBars, thick](axis cs:0.6,\hybridOverheadBatchOneDLRMBottomPZero)--(axis cs:0.6,\hybridOverheadBatchOneDLRMBottomPOneHundred);

\path [draw=\colorErrorBars, thick](axis cs:1.3,\threadOverheadBatchOneDLRMTopPZero)--(axis cs:1.3,\threadOverheadBatchOneDLRMTopPOneHundred);
\path [draw=\colorErrorBars, thick](axis cs:1.6,\globalOverheadBatchOneDLRMTopPZero)--(axis cs:1.6,\globalOverheadBatchOneDLRMTopPOneHundred);
\path [draw=\colorErrorBars, thick](axis cs:1.9,\hybridOverheadBatchOneDLRMTopPZero)--(axis cs:1.9,\hybridOverheadBatchOneDLRMTopPOneHundred);

\path [draw=\colorErrorBars, thick](axis cs:2.6,\threadOverheadBatchTwentyFourtyEightDLRMBottomPZero)--(axis cs:2.6,\threadOverheadBatchTwentyFourtyEightDLRMBottomPOneHundred);
\path [draw=\colorErrorBars, thick](axis cs:2.9,\globalOverheadBatchTwentyFourtyEightDLRMBottomPZero)--(axis cs:2.9,\globalOverheadBatchTwentyFourtyEightDLRMBottomPOneHundred);
\path [draw=\colorErrorBars, thick](axis cs:3.2,\hybridOverheadBatchTwentyFourtyEightDLRMBottomPZero)--(axis cs:3.2,\hybridOverheadBatchTwentyFourtyEightDLRMBottomPOneHundred);

\path [draw=\colorErrorBars, thick](axis cs:3.9,\threadOverheadBatchTwentyFourtyEightDLRMTopPZero)--(axis cs:3.9,\threadOverheadBatchTwentyFourtyEightDLRMTopPOneHundred);
\path [draw=\colorErrorBars, thick](axis cs:4.2,\globalOverheadBatchTwentyFourtyEightDLRMTopPZero)--(axis cs:4.2,\globalOverheadBatchTwentyFourtyEightDLRMTopPOneHundred);
\path [draw=\colorErrorBars, thick](axis cs:4.5,\hybridOverheadBatchTwentyFourtyEightDLRMTopPZero)--(axis cs:4.5,\hybridOverheadBatchTwentyFourtyEightDLRMTopPOneHundred);

\end{axis}

\end{tikzpicture}

%% file: figures/eval_conv_noscope.tex
\begin{tikzpicture}[]
\pgfplotsset{label style={font=\footnotesize}, 
             tick label style={font=\footnotesize},
             legend style={font=\footnotesize},
             every non boxed x axis/.append style={x axis line style=-},
             every non boxed y axis/.append style={y axis line style=-},
             axis lines=left,
             /pgfplots/ybar legend/.style={
                /pgfplots/legend image code/.code={%
                    \draw[##1,/tikz/.cd,yshift=-0.25em]
                    (0cm,0cm) rectangle (12pt,6pt);
                },
            },
        }

\begin{axis}[
height=\figureheight,
legend cell align={left},
legend columns=3,
legend style={at={(0.5,1.33)}, anchor=north, draw=none, /tikz/every even column/.append style={column sep=0.4cm}},
tick align=outside,
tick label style={/pgf/number format/assume math mode},
tick pos=left,
width=\figurewidth,
x grid style={white!69.01960784313725!black},
xmin=-0.41, xmax=4.81,
xticklabel style={align=center},
xtick={0.3,1.6,2.9,4.2},%
xticklabels={
{\shortNameNoscopeCoral\\(\aiBatchSixtyFourNoscopeCoral)},
{\shortNameNoscopeRoundabout\\(\aiBatchSixtyFourNoscopeRoundabout)},
{\shortNameNoscopeTaipei\\(\aiBatchSixtyFourNoscopeTaipei)},
{\shortNameNoscopeAmsterdam\\(\aiBatchSixtyFourNoscopeAmsterdam)}},
xlabel style={align=center,at={(axis description cs:0.5,-.2)}},
y grid style={gray, opacity=0.3},
ylabel style={at={(axis description cs:0.02,.5)}},
ymajorgrids,
ylabel style={align=left},
ylabel={\SlowdownMetric (\%)},
ymin=0, ymax=25
]

\addlegendimage{ybar,ybar legend,draw=black,fill=color_blue_3_2};
\addlegendentry{\ThreadABFT}

\addlegendimage{ybar,ybar legend,draw=black,fill=color_blue_3_1};
\addlegendentry{\GlobalABFT}

\addlegendimage{ybar,ybar legend,draw=black,fill=color_blue_3_0};
\addlegendentry{\HybridABFT}

\draw[draw=black,fill=color_blue_3_2] (axis cs:-0.15,0) rectangle (axis cs:0.15,\threadOverheadBatchSixtyFourNoscopeCoral);
\draw[draw=black,fill=color_blue_3_1] (axis cs:0.15,0) rectangle (axis cs:0.45,\globalOverheadBatchSixtyFourNoscopeCoral);
\draw[draw=black,fill=color_blue_3_0] (axis cs:0.45,0) rectangle (axis cs:0.75,\hybridOverheadBatchSixtyFourNoscopeCoral);

\draw[draw=black,fill=color_blue_3_2] (axis cs:1.15,0) rectangle (axis cs:1.45,\threadOverheadBatchSixtyFourNoscopeRoundabout);
\draw[draw=black,fill=color_blue_3_1] (axis cs:1.45,0) rectangle (axis cs:1.75,\globalOverheadBatchSixtyFourNoscopeRoundabout);
\draw[draw=black,fill=color_blue_3_0] (axis cs:1.75,0) rectangle (axis cs:2.05,\hybridOverheadBatchSixtyFourNoscopeRoundabout);

\draw[draw=black,fill=color_blue_3_2] (axis cs:2.45,0) rectangle (axis cs:2.75,\threadOverheadBatchSixtyFourNoscopeTaipei);
\draw[draw=black,fill=color_blue_3_1] (axis cs:2.75,0) rectangle (axis cs:3.05,\globalOverheadBatchSixtyFourNoscopeTaipei);
\draw[draw=black,fill=color_blue_3_0] (axis cs:3.05,0) rectangle (axis cs:3.35,\hybridOverheadBatchSixtyFourNoscopeTaipei);

\draw[draw=black,fill=color_blue_3_2] (axis cs:3.75,0) rectangle (axis cs:4.05,\threadOverheadBatchSixtyFourNoscopeAmsterdam);
\draw[draw=black,fill=color_blue_3_1] (axis cs:4.05,0) rectangle (axis cs:4.35,\globalOverheadBatchSixtyFourNoscopeAmsterdam);
\draw[draw=black,fill=color_blue_3_0] (axis cs:4.35,0) rectangle (axis cs:4.65,\hybridOverheadBatchSixtyFourNoscopeAmsterdam);

\path [draw=\colorErrorBars, thick](axis cs:0,\threadOverheadBatchSixtyFourNoscopeCoralPZero)--(axis cs:0,\threadOverheadBatchSixtyFourNoscopeCoralPOneHundred);
\path [draw=\colorErrorBars, thick](axis cs:0.3,\globalOverheadBatchSixtyFourNoscopeCoralPZero)--(axis cs:0.3,\globalOverheadBatchSixtyFourNoscopeCoralPOneHundred);
\path [draw=\colorErrorBars, thick](axis cs:0.6,\hybridOverheadBatchSixtyFourNoscopeCoralPZero)--(axis cs:0.6,\hybridOverheadBatchSixtyFourNoscopeCoralPOneHundred);

\path [draw=\colorErrorBars, thick](axis cs:1.3,\threadOverheadBatchSixtyFourNoscopeRoundaboutPZero)--(axis cs:1.3,\threadOverheadBatchSixtyFourNoscopeRoundaboutPOneHundred);
\path [draw=\colorErrorBars, thick](axis cs:1.6,\globalOverheadBatchSixtyFourNoscopeRoundaboutPZero)--(axis cs:1.6,\globalOverheadBatchSixtyFourNoscopeRoundaboutPOneHundred);
\path [draw=\colorErrorBars, thick](axis cs:1.9,\hybridOverheadBatchSixtyFourNoscopeRoundaboutPZero)--(axis cs:1.9,\hybridOverheadBatchSixtyFourNoscopeRoundaboutPOneHundred);

\path [draw=\colorErrorBars, thick](axis cs:2.6,\threadOverheadBatchSixtyFourNoscopeTaipeiPZero)--(axis cs:2.6,\threadOverheadBatchSixtyFourNoscopeTaipeiPOneHundred);
\path [draw=\colorErrorBars, thick](axis cs:2.9,\globalOverheadBatchSixtyFourNoscopeTaipeiPZero)--(axis cs:2.9,\globalOverheadBatchSixtyFourNoscopeTaipeiPOneHundred);
\path [draw=\colorErrorBars, thick](axis cs:3.2,\hybridOverheadBatchSixtyFourNoscopeTaipeiPZero)--(axis cs:3.2,\hybridOverheadBatchSixtyFourNoscopeTaipeiPOneHundred);

\path [draw=\colorErrorBars, thick](axis cs:3.9,\threadOverheadBatchSixtyFourNoscopeAmsterdamPZero)--(axis cs:3.9,\threadOverheadBatchSixtyFourNoscopeAmsterdamPOneHundred);
\path [draw=\colorErrorBars, thick](axis cs:4.2,\globalOverheadBatchSixtyFourNoscopeAmsterdamPZero)--(axis cs:4.2,\globalOverheadBatchSixtyFourNoscopeAmsterdamPOneHundred);
\path [draw=\colorErrorBars, thick](axis cs:4.5,\hybridOverheadBatchSixtyFourNoscopeAmsterdamPZero)--(axis cs:4.5,\hybridOverheadBatchSixtyFourNoscopeAmsterdamPOneHundred);

\end{axis}

\end{tikzpicture}

%% file: figures/eval_square.tex
\begin{tikzpicture}[]
\pgfplotsset{label style={font=\footnotesize}, 
             tick label style={font=\footnotesize},
             legend style={font=\footnotesize},
             every non boxed x axis/.append style={x axis line style=-},
             every non boxed y axis/.append style={y axis line style=-},
             axis lines=left,
             /pgfplots/ybar legend/.style={
                /pgfplots/legend image code/.code={%
                    \draw[##1,/tikz/.cd,yshift=-0.25em]
                    (0cm,0cm) rectangle (12pt,6pt);
                },
            },
        }

\begin{axis}[
height=\figureheight,
legend cell align={left},
legend columns=2,
legend style={at={(0.5,1.6)}, anchor=north, draw=none, /tikz/every even column/.append style={column sep=0.4cm}},
tick align=outside,
tick label style={/pgf/number format/assume math mode},
tick pos=left,
width=\figurewidth,
x grid style={white!69.01960784313725!black},
xmin=-0.41, xmax=10.91,
xticklabel style={align=center},
xtick={0.45,2.05,3.65,5.25,6.85,8.45,10.05},
xticklabels={{32\\(10.7)},{64\\(21.3)},{128\\(42.7)},{256\\(85.3)},{512\\(170.7)},{1024\\(341.3)},{2048\\(682.7)}},
xlabel style={align=center,at={(axis description cs:0.5,-.2)}},
xlabel={\Matmatmult size ($\gemmM = \gemmN = \gemmK$)},
y grid style={gray, opacity=0.3},
ylabel style={at={(axis description cs:0.05,.5)}},
ymajorgrids,
ylabel style={align=left},
ylabel={\SlowdownMetric (\%)},
ymin=0, ymax=35
]

\addlegendimage{ybar,ybar legend,draw=black,fill=color_blue_3_2};
\addlegendentry{\ThreadABFT (one-sided)}

\addlegendimage{ybar,ybar legend,draw=black,fill=color_blue_3_1};
\addlegendentry{Thread-level ABFT (two-sided)}

\addlegendimage{ybar,ybar legend,draw=black,fill=color_blue_3_0};
\addlegendentry{Thread-level replication}

\addlegendimage{ybar,ybar legend,draw=black,fill=color_red_2_1};
\addlegendentry{\GlobalABFT}

\draw[draw=black,fill=color_blue_3_2] (axis cs:-0.150,0) rectangle (axis cs:0.150,8);
\draw[draw=black,fill=color_blue_3_1] (axis cs:0.150,0) rectangle (axis cs:0.450,6);
\draw[draw=black,fill=color_blue_3_0] (axis cs:0.450,0) rectangle (axis cs:0.750,10);
\draw[draw=black,fill=color_red_2_1] (axis cs:0.750,0) rectangle (axis cs:1.050,27);

\draw[draw=black,fill=color_blue_3_2] (axis cs:1.450,0) rectangle (axis cs:1.750,4);
\draw[draw=black,fill=color_blue_3_1] (axis cs:1.750,0) rectangle (axis cs:2.050,12);
\draw[draw=black,fill=color_blue_3_0] (axis cs:2.050,0) rectangle (axis cs:2.350,8);
\draw[draw=black,fill=color_red_2_1] (axis cs:2.350,0) rectangle (axis cs:2.650,26);

\draw[draw=black,fill=color_blue_3_2] (axis cs:3.050,0) rectangle (axis cs:3.350,4);
\draw[draw=black,fill=color_blue_3_1] (axis cs:3.350,0) rectangle (axis cs:3.650,11);
\draw[draw=black,fill=color_blue_3_0] (axis cs:3.650,0) rectangle (axis cs:3.950,10);
\draw[draw=black,fill=color_red_2_1] (axis cs:3.950,0) rectangle (axis cs:4.250,21);

\draw[draw=black,fill=color_blue_3_2] (axis cs:4.650,0) rectangle (axis cs:4.950,4);
\draw[draw=black,fill=color_blue_3_1] (axis cs:4.950,0) rectangle (axis cs:5.250,11);
\draw[draw=black,fill=color_blue_3_0] (axis cs:5.250,0) rectangle (axis cs:5.550,9);
\draw[draw=black,fill=color_red_2_1] (axis cs:5.550,0) rectangle (axis cs:5.850,15);

\draw[draw=black,fill=color_blue_3_2] (axis cs:6.250,0) rectangle (axis cs:6.550,2);
\draw[draw=black,fill=color_blue_3_1] (axis cs:6.550,0) rectangle (axis cs:6.850,4);
\draw[draw=black,fill=color_blue_3_0] (axis cs:6.850,0) rectangle (axis cs:7.150,13);
\draw[draw=black,fill=color_red_2_1] (axis cs:7.150,0) rectangle (axis cs:7.450,8);

\draw[draw=black,fill=color_blue_3_2] (axis cs:7.850,0) rectangle (axis cs:8.150,13);
\draw[draw=black,fill=color_blue_3_1] (axis cs:8.150,0) rectangle (axis cs:8.450,20);
\draw[draw=black,fill=color_blue_3_0] (axis cs:8.450,0) rectangle (axis cs:8.750,77);
\draw[draw=black,fill=color_red_2_1] (axis cs:8.750,0) rectangle (axis cs:9.050,3);

\draw[draw=black,fill=color_blue_3_2] (axis cs:9.450,0) rectangle (axis cs:9.750,14);
\draw[draw=black,fill=color_blue_3_1] (axis cs:9.750,0) rectangle (axis cs:10.050,18);
\draw[draw=black,fill=color_blue_3_0] (axis cs:10.050,0) rectangle (axis cs:10.350,75);
\draw[draw=black,fill=color_red_2_1] (axis cs:10.350,0) rectangle (axis cs:10.650,1);

\addplot [thick, black, dashed]
table {%
7.65 0
7.65 35
};

\path [draw=\colorErrorBars, thick](axis cs:0,5.0841862)--(axis cs:0,10.36645758);
\path [draw=\colorErrorBars, thick](axis cs:0.3,5.275250227)--(axis cs:0.3,6.619654231);
\path [draw=\colorErrorBars, thick](axis cs:0.6,8.534675615)--(axis cs:0.6,12.11409396);
\path [draw=\colorErrorBarsGlobal, thick](axis cs:0.9,25.21029464)--(axis cs:0.9,28.83655175);

\path [draw=\colorErrorBars, thick](axis cs:1.6,3.102386451)--(axis cs:1.6,4.911470362);
\path [draw=\colorErrorBars, thick](axis cs:1.9,10.65485362)--(axis cs:1.9,13.18567026);
\path [draw=\colorErrorBars, thick](axis cs:2.2,6.694987256)--(axis cs:2.2,8.589634664);
\path [draw=\colorErrorBarsGlobal, thick](axis cs:2.5,24.85150742)--(axis cs:2.5,26.55767212);

\path [draw=\colorErrorBars, thick](axis cs:3.2,3.6431749)--(axis cs:3.2,4.720206041);
\path [draw=\colorErrorBars, thick](axis cs:3.5,10.64172535)--(axis cs:3.5,11.27112676);
\path [draw=\colorErrorBars, thick](axis cs:3.8,9.268587548)--(axis cs:3.8,10.43723554);
\path [draw=\colorErrorBarsGlobal, thick](axis cs:4.1,20.7064233)--(axis cs:4.1,21.4866936);

\path [draw=\colorErrorBars, thick](axis cs:4.8,3.717948718)--(axis cs:4.8,4.508547009);
\path [draw=\colorErrorBars, thick](axis cs:5.1,10.79932293)--(axis cs:5.1,11.56479218);
\path [draw=\colorErrorBars, thick](axis cs:5.4,8.641686183)--(axis cs:5.4,9.295081967);
\path [draw=\colorErrorBarsGlobal, thick](axis cs:5.7,14.53118792)--(axis cs:5.7,15.39531188);

\path [draw=\colorErrorBars, thick](axis cs:6.4,1.742996888)--(axis cs:6.4,2.178746109);
\path [draw=\colorErrorBars, thick](axis cs:6.7,3.683342496)--(axis cs:6.7,4.178119846);
\path [draw=\colorErrorBars, thick](axis cs:7.0,12.8220571)--(axis cs:7.0,13.1474384);
\path [draw=\colorErrorBarsGlobal, thick](axis cs:7.3,7.81669239)--(axis cs:7.3,8.187106732);

\path [draw=\colorErrorBars, thick](axis cs:8.0,12.94584611)--(axis cs:8.0,13.02758783);
\path [draw=\colorErrorBars, thick](axis cs:8.3,19.95848311)--(axis cs:8.3,20.03949167);
\path [draw=\colorErrorBars, thick](axis cs:8.6,76.92044427)--(axis cs:8.6,77.07955573);
\path [draw=\colorErrorBarsGlobal, thick](axis cs:8.9,2.960048426)--(axis cs:8.9,3.032687651);

\path [draw=\colorErrorBars, thick](axis cs:9.6,12.84590467)--(axis cs:9.6,16.02036094);
\path [draw=\colorErrorBars, thick](axis cs:9.9,14.4734245)--(axis cs:9.9,22.57718878);
\path [draw=\colorErrorBars, thick](axis cs:10.2,71.5993972)--(axis cs:10.2,78.16472284);
\path [draw=\colorErrorBarsGlobal, thick](axis cs:10.5,0.1931567329)--(axis cs:10.5,2.075055188);

\end{axis}

\end{tikzpicture}

%% file: discussion.tex
\section{Discussion} \label{sec:discussion}
\cutcandidate{
\subsection{Additional opportunities for adaptive \abft} \label{sec:discussion:thread}
Our goal in \hybridABFT is to adapt the \abft scheme used depending on the resource bottlenecks of a given \layer so as to reduce execution-time overhead. However, there are additional differences between \threadABFT and \globalABFT that can be exploited to adapt \abft to other per-\layer properties. 

For example, \threadABFT and \globalABFT provide different fault tolerance guarantees. Recall that both \globalABFT and \threadABFT are capable of detecting a single faulty output resulting in the \matmatmult over which checksums are generated. Since \globalABFT generates checksums across the entire \matmatmult performed by the \kernel, it can detect a single fault across the entire \kernel-level \matmatmult. In contrast, since \threadABFT generates checksums across small, thread-level \matmatmults, it can detect one fault per thread in the \kernel, resulting in higher overall fault tolerance guarantees. As prior work has illustrated that different \layers in \nns are more susceptible to faults than others~\cite{li2017understanding,schorn2018accurate}, one might also consider varying the approach to \abft used depending on the fault tolerance requirement of a given \layer.} 

\subsection{\HybridABFT beyond \nns} \label{sec:discussion:nn}
While we have focused the design of \hybridABFT for imparting fault tolerance to \nn inference, \hybridABFT is applicable to general \matmatmult problems as well. We consider this particularly important as more traditional HPC applications begin exploring the use of \nn hardware accelerators, such as \tensorcores~\cite{haidar2018harnessing,dakkak2019accelerating,feng2021egemm}, and as \nn hardware accelerators begin to add support for double- and single-precision floating point arithmetic~\cite{nvidia-ampere}, which are typically used in HPC applications.

\cutcandidate{\subsection{\HybridABFT beyond GPUs} \label{sec:discussion:gpu}
While our discussion in \Section\ref{sec:opp} motivating the mix of compute- and bandwidth-bound \linearlayers in \nns was centered on trends in GPU hardware, \hybridABFT can benefit other \nn accelerators as well. For example, TPUv1 has a high \cmr of over 1000~\cite{jouppi2017datacenter}. Many \linearlayers in \nns will thus likely underutilize the computational units on such accelerators, requiring adaptive, arithmetic-intensity-driven \abft selection. 

As we have shown, key to being able to exploit compute underutilization at a fine granularity in such accelerators is the ability to embed redundant execution alongside the \matmatmult itself. This requires hardware vendors to enable accelerators to be programmed at a low enough level for redundant execution to, for example, share loads with the original \matmatmult. We are hopeful that future \nn accelerators will provide software capabilities similar to GPUs to allow low-level control.}

\revised{\subsection{Mathematical models for \abft}
In this work, \hybridABFT leverages empirical profiling to make the final decision of which \abft scheme to use for a given layer of a \nn. An alternative is to leverage analytical models of \abft with assumptions about compute and memory bandwidth to analytically determine which approach to \abft will likely result in lower execution-time overhead. \HybridABFT could also leverage such models. We have chosen to use empirical profiling because this is the common practice used in optimizing \nns for inference by popular frameworks. Regardless of whether empirical profiling or analytical modeling is used, the core insights driving \hybridABFT will remain relevant: layers with low \aiShort relative to a GPU's \cmr are likely to benefit from \threadABFT, while layers with high \aiShort relative to a GPU's \cmr are likely to benefit from \globalABFT.

\subsection{Input-size-dependent optimization}
As noted in \Section\ref{sec:opp:ai}, the size of the input to a \nn affects its \ai, and thus the selection made by \hybridABFT. Thus, one might wonder whether \hybridABFT may be suboptimal if a deployment experiences changes in the size of inputs. This, however, is not a major concern because it is rare to have dynamically-sized inputs at inference time due to the common preprocessing step for \nn inference of resizing inputs to a fixed format. The input-size-dependent heterogeneity of \aiShort described in \Section\ref{sec:opp:ai} will exist across deployments, but is unlikely within a single deployment. If multiple input sizes are expected within a deployment, one can handle this easily by performing separate \abft selections for multiple input sizes and choosing among these at inference time depending on the size of an input. 

}

%% file: related.tex
\section{Related work} \label{sec:related}
\textbf{Fault tolerance in general programs.} There are various techniques for tolerating soft-error-induced faults in general programs:

One approach is hardware-based fault tolerance, such as through using ECC in memory, and radiation-hardened or redundant processing units~\cite{dell1997white,bartlett2004commercial,sullivan2018swapcodes}. While certain approaches to hardware-based fault tolerance are widely used, such as ECC-protected memory subsystems, hardware protection for processing units is less widely used due to its high overhead. Furthermore, hardware-based fault tolerance is inflexible to changes in the required fault tolerance of applications or the fault rate of operating environments. Thus, we focus on software-based fault tolerance.

Software-based fault tolerance for general programs is typically achieved through techniques like instruction duplication~\cite{reis2005swift,mahmoud2018optimizing}, replication of threads/warps~\cite{dimitrov2009understanding,jeon2012warped,wadden2014real,yang2021enabling}, and compiler-driven re-execution~\cite{liu2016compiler,kim2020compiler}. In contrast, we focus on using application-level features of \nn inference to reduce the overhead of fault detection. %

\textbf{Fault tolerance in \nns.}
Recent works have illustrated the potentially-catastrophic effects of \ses on \nns through fault injection tools~\cite{li2017understanding,chen2019binfi,chen2020tensorfi,mahmoud2020pytorchfi} and neutron beam experiments~\cite{dos2019reliability}. This has spurred many approaches for fault tolerance in \nns, such as leveraging the ``inherent robustness'' of \nns~\cite{torres2017fault,zhang2018thundervolt,ozen2020concurrent}, training \nns to tolerate faults~\cite{koppula2019eden}, anomalous activation suppression~\cite{ozen2020just,chen2021ranger}, selective feature hardening~\cite{mahmoud2020hardnn}, and learning to detect faults~\cite{schorn2018efficient,li2020deepdyve,schorn2020facer}. Our focus in this work is on leveraging \abft to detect errors in \nn inference. Compared to the approaches listed above, \abft provides clearer fault-tolerance guarantees and does not require retraining a \nn or understanding its behavior.

\textbf{\abft for \nns.} Due to the heavy use of linear algebra in \nns, \abft is a natural fit for fault tolerance in \nns, and a number of recent works have explored  using \abft for \nns~\cite{dutta2019codenet,ozen2019sanity,zhao2021ft,hari2021making,li2021efficient}. Ozen et al.~\cite{ozen2019sanity} leverage \abft to protect convolutional and fully-connected \layers and propose integration of \abft into a systolic array architecture.  Zhao et al.~\cite{zhao2021ft} propose a systematic workflow of \abft checks for \cnns to provide a high degree of protection against faults with low execution-time overhead on CPUs. Li et al.~\cite{li2021efficient} propose optimizations for \abft in low-bitwidth \dlrm inference on CPUs. Most closely related to our work is the work of \harietal, which proposes the optimized \globalABFT scheme for GPUs (described in \Section\ref{sec:background:abft_nn}), and which forms a component of our proposed \hybridABFT. \HybridABFT complements the work of \harietal with \abft schemes well-suited for bandwidth-bound \linearlayers, and by adaptively selecting between the two, using \aiShort as a guide.

Compared to these works, the present work is unique in multiple aspects. First, the works listed above all focus on employing a single \abft scheme across all \linearlayers of a \nn. In contrast, we illustrate that different \linearlayers within a \nn have varying resource bottlenecks that benefit from per-\layer-optimization in the proposed \hybridABFT. Second, to the best of our knowledge, our work is the first to analyze the growing bottleneck of memory bandwidth for \nn inference in the context of exploiting it for efficient redundant execution. Careful analysis of this trend lends itself to developing optimizations that have been overlooked by prior works. Finally, to the best of our knowledge, this work presents the first thread-level approach to \abft for \nns on GPUs. 

\textbf{\abft in other domains.} 
\abft has been widely studied for imparting fault tolerance to linear algebra~\cite{huang1984algorithm,bosilca2009algorithm,braun2014abft,wu2016towards,zamani2019greenmm}, iterative methods~\cite{chen2013online,chen2016online}, and other applications~\cite{li2019ft}. Our work differs from these works in its focus on the specific characteristics of \nns and GPUs, and its adaptive, intensity-guided approach of selecting \abft schemes based on the resource bottlenecks of the problem. 

Smith et al.~\cite{smith2015toward} investigated fusing \abft operations alongside \matmatmults on CPUs. While similar to the approach to \threadABFT that we consider as part of \hybridABFT, the techniques employed by Smith et al.~\cite{smith2015toward} differ in that they do not perform \abft at the level of the smallest unit of the parallel subproblem in the \matmatmult. Thus, this approach generates checksums collaboratively across CPU threads (although not globally), which requires additional loads and stores, albeit, at higher levels of the memory hierarchy. In contrast, we leverage \threadABFT specifically for bandwidth-bound \linearlayers in \nns on GPUs, and thus avoid performing any additional loads and stores (which would compete for the \layer's bottleneck resource). This results in \threadABFT performing \abft at the smallest parallel sub-\matmatmult solved (GPU thread level), requiring no coordination between threads. Furthermore, \hybridABFT takes an adaptive approach to \abft based on the resource bottleneck of a given \matmatmult, whereas Smith et al.~\cite{smith2015toward} use a one-size-fits-all approach.

Concurrent work with ours, FT-BLAS~\cite{zhai2021ft}, proposes to choose between replication and \abft in BLAS routines on CPUs depending on the BLAS level of an operation. Specifically, FT-BLAS~\cite{zhai2021ft} uses replication for operations in Levels~1 and 2 (vector-vector and matrix-vector operations), and \abft for those in Level~3  (matrix-matrix operations). However, for a given operation in a BLAS level (\eg for all matrix-matrix multiplications), FT-BLAS~\cite{zhai2021ft} takes a one-size-fits-all approach. In contrast, we show that the unique characteristics of \nn inference on GPUs lead to \textit{\nns containing a mix of compute- and bandwidth-bound matrix-matrix multiplications}, rendering one-size-fits-all approaches inefficient.  {Moreover, as shown in \Section\ref{sec:eval:square}, leveraging replication even for bandwidth-bound \matmatmults used in \nns on GPUs can lead to significant overhead, motivating \hybridABFT's approach of selecting between various \abft schemes for each \matmult.} 

%% file: conclusion.tex
\section{Conclusion}
We present \hybridABFT, a new approach to \abft for \nn inference on GPUs that optimizes for the specific resource bottlenecks of individual layers of a \nn. Through analysis of trends in \nn design and GPU hardware, we present a case for a growing trend of compute underutilization in \nn inference on GPUs, opening new opportunities for efficient redundant execution. However current approaches to \abft for \nns are unable to exploit such fine-grained compute underutilization. We first carefully investigate a \threadABFT scheme to exploit such opportunities in bandwidth-bound \linearlayers of \nns on GPUs, complementing the use of traditional approaches to \abft for compute-bound \layers. \HybridABFT then adaptively selects among these \abft schemes in a per-layer, arithmetic-intensity-driven fashion. This enables \hybridABFT to reduce execution-time overhead by 1.09--5.3$\times$ across a number of popular and emerging \nns. \HybridABFT shows the promise of arithmetic-intensity-driven fault tolerance for current and future \nns. Finally, as \hybridABFT protects general \matmatmultsShort, this approach may usher more efficient fault tolerance for broader HPC applications that are beginning to target \nn hardware accelerators~\cite{haidar2018harnessing,dakkak2019accelerating,feng2021egemm}. %

\section*{Acknowledgements}
This work was funded in part by a National Science Foundation Graduate Research Fellowship (DGE-1745016 and
DGE-1252522), in part by Amazon Web Services, and in part by the AIDA project (POCI-01-0247-FEDER-045907) co-financed by the European Regional Development Fund through the Operational Program for Competitiveness and Internationalisation 2020. %